\documentclass[a4paper,english,final,twocolumn,3p,sort&compress]{elsarticle}
\usepackage[T1]{fontenc}
\usepackage[utf8]{inputenc}
\usepackage{color}
\usepackage{babel}
\usepackage{array}
\usepackage{prettyref}
\usepackage{booktabs}
\usepackage{textcomp}
\usepackage{url}
\usepackage{multirow}
\usepackage{amsthm}
\usepackage{amsmath}
\usepackage{amssymb}
\usepackage{graphicx}
\usepackage[unicode=true,
 bookmarks=true,bookmarksnumbered=false,bookmarksopen=false,
 breaklinks=false,pdfborder={0 0 1},backref=false,colorlinks=false]
 {hyperref}
\usepackage{breakurl}

\makeatletter

\newcommand{\lyxmathsym}[1]{\ifmmode\begingroup\def\b@ld{bold}
  \text{\ifx\math@version\b@ld\bfseries\fi#1}\endgroup\else#1\fi}

\providecommand{\tabularnewline}{\\}

\newenvironment{lyxlist}[1]
{\begin{list}{}
{\settowidth{\labelwidth}{#1}
 \setlength{\leftmargin}{\labelwidth}
 \addtolength{\leftmargin}{\labelsep}
 }}
{\end{list}}

\@ifundefined{date}{}{\date{}}
\journal{Chemical Engineering \& Processing: Process Intensification}
\usepackage[labelfont={bf},nooneline]{caption}
\usepackage[labelformat=simple]{subfig}

\AtBeginDocument{

}
\AtBeginDocument{\renewcommand{\ref}[1]{\mbox{\autoref{#1}}}}
\DeclareInputText{183}{\ifmmode\cdot\else\textperiodcentered\fi}

\@ifundefined{showcaptionsetup}{}{%
 \PassOptionsToPackage{caption=false}{subfig}}
\usepackage{subfig}
\makeatother

\begin{document}

\title{Molten salts database for energy applications}

\author[rvt]{R.~Serrano-López\corref{cor1}}

\ead{robertosl@ubu.es}

\author[rvt]{J.~Fradera}

\ead{jfradera@ubu.es}

\author[rvt]{S.~Cuesta-López\corref{cor2}}

\ead{scuesta@ubu.es}

\cortext[cor1]{Corresponding author. Tel: (+34) 947258907. Universidad de Burgos,
Spain. }

\cortext[cor2]{Principal corresponding author. Tel: (+34) 947259062. Universidad
de Burgos, Spain.}

\address[rvt]{Science and Technology Park. I+D+I Building. Room 63. Plaza Misael
Bañuelos s/n, 09001, Burgos (Spain) }
\begin{abstract}
The growing interest in energy applications of molten salts is justified
by several of their properties. Their possibilities of usage as a
coolant, heat transfer fluid or heat storage substrate, require thermo-hydrodynamic
refined calculations. Many researchers are using simulation techniques,
such as Computational Fluid Dynamics (CFD) for their projects or conceptual
designs. The aim of this work is providing a review of basic properties
(density, viscosity, thermal conductivity and heat capacity) of the
most common and referred salt mixtures. After checking data, tabulated
and graphical outputs are given in order to offer the most suitable
available values to be used as input parameters for other calculations
or simulations. The reviewed values show a general scattering in characterization,
mainly in thermal properties. This disagreement suggests that, in
several cases, new studies must be started (and even new measurement
techniques should be developed) to obtain accurate values.\end{abstract}
\begin{keyword}
properties \sep molten salt \sep CFD \sep CSP \sep coolants \sep
energy 
\end{keyword}
\maketitle

\section*{Nomenclature}

\subsection*{{\normalsize Symbols and Units}}
\begin{lyxlist}{00.00.0000}
\item [{\quad{}T}] Temperature, \textdegree{}K
\item [{\quad{}M.P.}] Melting point, \textdegree{}K
\item [{\quad{}v}] Velocity, m/s
\item [{\quad{}$\rho$}] Density, kg/m\textthreesuperior{}
\item [{\quad{}$\eta$}] Dynamic viscosity, Pa·s
\item [{\quad{}$\nu=\frac{\eta}{\rho}$}] Kinematic viscosity, m\texttwosuperior{}/s
\item [{\quad{}$\lambda$}] Thermal conductivity, W/(m·\textdegree{}K)
\item [{\quad{}$\mathit{C}p$}] Heat Capacity, J/(kg·\textdegree{}K)
\end{lyxlist}

\subsection*{{\normalsize Salt Mixtures}}

\subsubsection*{\textmd{\quad{}Fluorides}}
\begin{lyxlist}{00.00.0000}
\item [{\quad{}$2\mathrm{LiF-BeF_{2}}$}] FLiBe
\item [{\quad{}$\mathrm{LiF-NaF-KF}$}] FLiNaK
\item [{\quad{}$\mathrm{LiF-NaF-BeF_{2}}$}] FLiNaBe
\item [{\quad{}$\mathrm{NaF-NaBF_{4}}$}] Denoted by us as NaFNaB
\item [{\quad{}$\mathrm{KF-ZrF_{4}}$}] Denoted by us as FluZirK
\end{lyxlist}

\subsubsection*{\textmd{\quad{}Chlorides	}}
\begin{lyxlist}{00.00.0000}
\item [{\quad{}$\mathrm{KCl-MgCl_{2}}$}] Denoted by us as CloKMag
\end{lyxlist}

\subsubsection*{\textmd{\quad{}Nitrates}}
\begin{lyxlist}{00.00.0000}
\item [{\quad{}$\mathrm{NaNO_{3}-KNO_{3}}$}] Solar~Salt
\item [{\quad{}$\mathrm{NaNO_{3}-NaNO_{2}-KNO_{3}}$}] Hitec$^{\lyxmathsym{\textregistered}}$ 
\end{lyxlist}

\subsection*{{\normalsize Acronyms and Abbreviations}}
\begin{lyxlist}{00.00.0000}
\item [{\quad{}ACS}] American Ceramic Society
\item [{\quad{}AHTR}] Advanced High-Temperature Reactor
\item [{\quad{}ARE}] Aircraft Reactor Experiment
\item [{\quad{}CFD}] Computational Fluid Dynamics
\item [{\quad{}CSP}] Concentrated Solar Power
\item [{\quad{}CSPonD}] Concentrated Solar Power on Demand
\item [{\quad{}FHR}] Fluoride Salt Cooled High-Temperature Reactor
\item [{\quad{}HR}] Homogenueus Reactor
\item [{\quad{}HTS}] Heat Transfer Fluid
\item [{\quad{}HTX}] Heat Exchanger
\item [{\quad{}IHX}] Intermediate Heat Exchanger
\item [{\quad{}INL}] Idaho National Laboratory
\item [{\quad{}MSRE}] Molten Salt Reactor Experiment
\item [{\quad{}NREL}] National Renewable Energy Laboratory
\item [{\quad{}ORNL}] Oak Ridge National Laboratory
\item [{\quad{}TES}] Thermal Energy Storage
\end{lyxlist}

\section{Introduction}

Different heat transfer systems are being studied around the use of
molten salt technology as working fluid. The initial development of
salts features and use was carried out at Oak Ridge National Laboratory
(ORNL) for aircraft propulsion purposes~\citep{Renault}. The ORNL
team build demonstration reactors between 1950's and 1960's (ARE and
MSRE), and continued their research during several years in order
to design a Molten Salt Breeder Reactor \citep{Forsberg2002}. During
these years, a great amount of technical reports about salt candidates
and tests were written, revised and archived, including a review of
the final experimental successes \citep{Haubenreich}. The experience
collected in that decades has been the subject of a renewed interest,
due to potential advantages of liquid salt coolants in different ways~\citep{Renault,Le_Brun2007,Forsberg2005_2}:\label{List-advantages}
\begin{itemize}
\item high volumetric heat capacity,
\item high boiling point and low vapor pressure,
\item no undesirable chemical exothermic reactions between different zones
of energy plants and power cycle coolants (core, heat exchange loop),
\item optical transparency during inspection operations,
\item ability to dissolve actinides,
\item great insensitivity to radiations.
\end{itemize}
Molten salts also offer nuclear breeding capability by optimizing
mixture and giving a on-line processing to the fuel salt carrier~\citep{Forsberg2002},
and certain compositions can be used as neutron moderators (e.g. those
containing Be). The advantages of salt coolants enable them for being
used in heat transport loops, and would open new nuclear and non-nuclear
applications~\citep{Le_Brun2007,Forsberg2004,Gen_IV_2009,ALISIA_D-50}~\label{list-applications}: 
\begin{itemize}
\item fission power plants (both liquid or solid fuel, mainly in high-temperature
reactors),
\item fusion or hybrid reactors,
\item hydrogen production,
\item long distance heat transport,
\item nuclear fuel reprocessing,
\item chemical industry,
\item oil refineries,
\item shale oil processing, etc. 
\end{itemize}
There are also advanced studies about the use of fluids in the power
cycle, transport and heat storage in high-efficiency solar power facilities~\citep{Forsberg2007,Slocum2011,flueckiger2012second,flueckiger2011integrated}.
Even future nuclear-propelled space aircrafts are being now studied
around the same concepts~\citep{Eades2012}. Developments achieved
in every of these fields could be generalized for civil and industrial
uses in the future with the benefits of scale economy and standardization
of processes.

The present work intends to be a thorough review of the current knowledge
on molten salt properties, with the aim of giving to the researcher
a key to evaluate their behavior. These data may be of interest in
both experimental as numerical simulation tasks, such Computational
Fluid Dynamics (CFD), and can be applied for thermal exchange, storage
or piping systems~\citep{Nunes2003}.

\section{Background\label{sec:Background}}

As has already be mentioned, the initial studies about salt properties
were developed by scientists involved in chain reactions. Harold Urey,
a nuclear chemist, and Eugene Wigner, a theoretical physicist, both
implicated in the Manhattan Project, were the early promoters of the
Homogeneous Reactors (HR). First molten fluorides ideas came into
the chain reaction community by 1945~\citep{weinberg1997proto}.
Therefore, ORNL Chemical Technology Division was searching since the
late forties for a homogeneous liquid solution suitable for use as
a fuel for the aircraft reactor of the Aircraft Nuclear Propulsion
Project. A fluid able for being used at high temperature and low pressure
was needed, with the requirement of radiation stability and actinides
solubility~\citep{briant1957molten}. It is assumed that Edward Bettis
and Raymond Clare Briant were the persons who, years later, cut the
Gordian knot by suggesting again the use of fluorides with the Molten
Salt Reactor concept~\citep{macpherson1985molten,rosenthal2009account}. 

Although initial tests with HR were developed through sulphates, the
pressure of Cold War shoved US Atomic Energy Commission to start the
Aircraft Nuclear Propulsion Program~\citep{rosenthal2009account}.
In this case, new requirements were solved by the use of liquid alkali
fluorides instead of solid fuel rods. Several developments in hi-tech
materials and salt chemistry (e.g. dry fluorination) were added to
knowledge during this experiences. From the very beginning, structural
materials corrosion appeared as the most important issue to resolve~\citep{briant1957molten}.
Lithium and beryllium salts were very early considered to allow self-moderated
fuels~\citep{ellis1950aircraft}, and FLiNaK appeared simultaneously
in heat transfer studies~\citep{grele1954forced,hoffman1955fused}.
Using graphs and charts, Poppendinck~\citep{poppendick1952physical}
analyzed some useful properties for different coolants, including
molten salts.

During the forties and the fifties some reports were also written
for ceramic applications (see, e.g.,~\citep{hall1933phase,hall1938phase,hall1947phase,hall1949phase,levin1956phase,levin1959phase}).
Since 1964, there have been different compilations of phase equilibria
diagrams at the National Bureau of Standards (edited by ACS)~\citep{levin1964phase}.
This multi-volume work is continuously improved performing the cumulative
specialized ACerS-NIST Phase Equilibria database (ACerS-NIST Phase
Equilibria Diagrams), which is also on-line nowadays (\url{http://ceramics.org/publications-and-resources/phase-equilibria-diagrams}).

Ricci~\citep{ricci1958guide} and Thoma~\citep{thoma1959phase}
made the first efforts to compile specific available equilibria data
for different kind of chemical systems for nuclear applications. Maybe
the first review of fused salt properties was made by Grimes~\citep{lane1958fluid}
for nuclear reactors coolants (FLiBe, FLiNaK, and FLiNaBe). Grimes
included measured values and correlations for Melting Point (MP),
density ($\rho$), heat capacity ($Cp$) and viscosity ($\eta$) for
these salts.

During sixties, several achievements were made around the Molten Salt
Reactor concept, including the referred use of salts as coolants in
heat exchange systems. Most of the reports came from ORNL Divisions.
Blanke et al.~\citep{Blanke1956} studied Li and Be fluorides density
and viscosity. Grimes~\citep{Grimes1964} gave some values for FLiBe
{[}66-34 \%{]} at 854 \textdegree{}K ($\rho,$ $\mathit{Cp}$, $\eta,$
\emph{$\lambda$}). Thereafter, Grimes~\citep{Grimes1967} discussed
the requirements of coolant fluids for the MSBR providing values for
NaFNaB {[}4-96 \%{]} and FLiNaBe {[}5-56-42 \%, and 26-41-36 \%{]}
among other salt mixtures at 727 \textdegree{}K ($\rho,$ $\mathit{Cp}$,
$\eta,$ \emph{$\lambda$}$)$~\citep{Grimes1967}. 

Cantor~\citep{Cantor1965,Cantor_et_al1968,Cantor1969,cantor_1969,Cantor1973}
studies are also a classical reference for salt properties. Two mixtures
were listed in 1968~\citep{Cantor_et_al1968}, FLiBe {[}66-34 \%{]}
and NaFNaB {[}92-89 \%{]}, based in a critical review of own and other
measurements. Then FLiBe was again studied in 1969~\citep{cantor_1969},
giving density and viscosity correlations for different melt compositions.
Some fluoride mixtures (FLiBe, NaFNaB) were revised also by Cantor
in 1973~\citep{Cantor1973}.

McDuffie et al.~\citep{McDuffie1969} reviewed again coolant requirements
for heat exchange for nuclear technology purposes, summarizing physical
properties of different mixes of fluorides, chlorides, nitrates, and
fluoborates. Similar works were conducted by Sanders (1971)~\citep{Sanders1971}
(NaFNaB, FLiBe, FLiNaK, nitrates), and Kelmers et al. (1976)~\citep{Kelmers1976}
(NaFNaB, FLiBe). The use of low-cost molten salts as heat transfer
fluids and their potential for thermal energy storage were also early
discussed; Silverman and Engel made a review of Solar Salt and Hitec$^{\lyxmathsym{\textregistered}}$
 capabilities in 1977~\citep{Silverman1977}.

Janz started in 1968 a huge effort to compile a general database for
molten salts~\citep{janz1968,janz1969} relevant to energy storage~\citep{janz1978physical,janz1979physical,janz1981physical}.
Reports were published by the Office of Standard Reference Data at
the National Bureau of Standards (OSRD-NBS), including a comprehensive
compilation of eutectic compositions for salts. This series made a
critical review of measurement methods and correlations, and were
progressively written and published between 1968 and 1981. 

Janz also worked in cumulative results for the salts properties between
1972 and 1983 (the Molten Salts Standards Program)~\citep{Janz1972,Janz1974,Janz1975,Janz1977,Janz1979,Janz1980,Janz1983}.
This general database, subsequently reprinted and summarized in 1988~\citep{Janz1988},
is widely used today for any purpose due to its updated recommendations
and re-examined best values for making density and viscosity predictions~\citep{Fuller2003}.

Lately, an important assessment of some liquid salts was conducted
by Williams et al. in 2006 at ORNL \citep{williams2005chemical,williams2006assess12,williams2006assess69},
focusing their works in the Advanced High-Temperature Reactor (AHTR)
initial concept development (primary and secondary coolants). This
work included classical Janz, Cantor and ORNL entrances in most usual
salt mixtures for nuclear reactors. However, Be-containing salts were
in these case excluded because of their higher cost and toxicity.
These studies also advised about taking into account $\mathrm{LiCl-KCl-MgCl_{2}}$
due to its low-cost. Further, Williams included additional measurements
for relatively unexplored Zr and Al fluoride salts \citep{williams20066additional}. 

During these last years, Winconsin-Madison University has joined with
Shell Company to perform the on-line Molten Salts Database. This American
institution is currently working to maximize molten salts potential
in energy issues (in support of fusion reactor and Very-High-Temperature
Reactor (VHTR) concepts), with emphasis in FLiNaK and CloKMag mixtures\citep{Sabharwall2010}.
Additional efforts are being conducted in the Idaho National Laboratory
(INL)~ \citep{calderoni2010experimental,Sohal2010}.

Other institutions have funded knowledge on liquid salts behavior
and selection criteria, like the International Atomic Energy Agency
(IAEA)~\citep{Samuel2009}, the International Science and Technology
Center (ISTC) by the support of European Community~\citep{Zherebtsov2001},
and the European Commission through ALISIA (Assessment of LIquid Salts
for Innovative Applications)~\citep{ALISIA_D-50}.

\section{Selection of salts}

The advantages of molten salts as Heat Transfer Fluid (HTF) and Thermal
Storage System (TES) promise a great development during next decades.
The cost for the required volume of heat exchangers and pumps are
highly reduced by the use of liquid salts instead of other coolants
due to their higher volumetric heat capacity without the need of pressurizing.
It has been reported that melting points and heat capacities increase
in the following order: nitrates, chlorides, carbonates, and fluorides
\citep{hoshi2005screening}. In any case, fluid salts provide the
potential for improved heat transfer and reduced pumping powers and
volume of the heat exchanger compared with helium. Molten salts have
a 25\% higher volumetric heat capacity than pressurized water, nearly
five-times that of liquid sodium~\citep{LeBlanc2010}, and more than
twice than lead or lead-bismuth eutectic~\citep{Nuclear-Agency-EnergyOECD2007,Forsberg2004}.

Long term corrosion, compatibility with available structural materials,
potential toxicity and final costs are key research issues. In some
cases, stability requirements reduce the number of chance possibilities;
in others, avoid freezing maybe the critical issue~\citep{siegel2012thermal}.
So the salt must be properly selected in order to agree the particular
conditions of use.

Be mixtures require very special and expensive handling efforts due
to toxicity, and so has been discarded from some reports~\citep{williams2006assess69}.
For this reason FLiNaK has emerged, in several cases, as main alternative
because of its low toxicity, excellent heat transfer properties, and
chemical properties similar to those of FLiBe~\citep{sohal2010conceptual}.
In nuclear applications, purity of Li is an additional requirement
with primary coolants (at least 99,995\% of ${{}^7}$Li is needed
to avoid decrease reactivity feedback, due to tritium generation by
the neutron absorption of $^{\mathrm{6}}$Li isotope), and even with
secondary coolants in case of a leak at Heat Exchanger (HTX) loop.
For this reason other possibilities (other than FLiBe and FLiNaK)
are now being explored for the latest versions of VHTR Intermediate
Heat Exchanger (IHX), e.g. FluZirK liquid salt, wich has a relatively
low toxicity, and does not include Li~\citep{holcomb2012current}.
Williams discussed the influence of the price of the components with
different salt mixtures~\citep{williams2006assess69}. His conclusions
determined that magnesium chlorides are the least expensive of all,
while fluorides, fluoroborates and Li-containing mixtures increase
the price of the coolant (Williams excluded Be because of the same
issues explained above). Hence, with a multi-criteria analysis including
technical and economic factors, an additional study of $\mathrm{MgCl_{2}}$-containing
salts should be recommended (despite of inferior heat-transfer metrics
of chlorides). The ideas previously summarized have suggested both
FluZirK and CloKMag as possible working fluids in heat transfer piping
systems. Several studies have been focused towards the behavior of
the new potential fluids~\citep{anderson2012molten,Ambrosek2011a,Glazoff2012,holcomb2011core}. 

Zirconium fluorides were used in the past given the fact of their
high solubility for actinides and oxygen getter. In addition, Zr has
been proposed to mitigate materials corrosion by controlling the red-ox
potential of the salts, but the activation products complicate their
ability to be used as primary coolants~\citep{Sabharwall2010}. Heavy
halide salts (bromides, iodides) have poor heat transfer metrics~\citep{williams2006assess69},
but chlorides, as well as nitrites and nitrates may be useful respect
to raw materials cost. Oxygen-containing salts, such as nitrite-nitrate
salt mixtures are not suitable as primary coolants in nuclear plants~\citep{ingersoll2006trade}.
Chlorides were avoided in the past due to corrosion, but there are
renewed expectatives around them in AHTR design.

New salt compositions are probably going to be tested for large scale
energy systems, and even nanofluids could emerge as future solutions
increasing $Cp$ of usual base fluids by the addition of nanoparticles
in low percentages~\citep{schuller2012molten}. However, after a
literature survey, classical mixtures appear in most of conceptual
and commercial designs of both nuclear and non nuclear applications.
For instance, recent reports~\citep{ALISIA_D-50,Samuel2009,Benes2012}
have reviewed most usual proposals for nuclear reactors, including
GEN-IV concepts.

Synergistic efforts in heat transport liquids are now being developed
for both molten and liquid salt concepts, especially taking into account
new possibilities in TES and Concentrated Solar power (CSP) technologies~\citep{Slocum2011,Denholm2012}.
The need of decarbonizing the energy mix current century has attracted
and renewed the attention to different studies started decades ago,
when nuclear and aerospace research was growing up~\citep{lacy1987selection,misra1987fluoride}.
In the other hand, first commercial CSP-TES systems are using common
compositions as working fluids, whose behavior is well known and have
more competitive cost. These salt mixtures take in advance the accrued
accomplishments since middle past century in heat transfer and storage,
looking for compromise between thermal efficiency and economy~\citep{siegel2012thermal}.
For CSP and TES, the raw materials cost of nitrogen-based salt-compositions
makes them currently the most competitive solution (Solar salt and
Hitec$^{\lyxmathsym{\textregistered}}$). Some Li-containing mixtures
and other additives are also been investigated and patented for future
development~\citep{Gomez2011,Li2013}. In this case nitrate-nitrites
are not only cheaper, but also much less corrosive than fluorides
or chlorides, while thermal capabilities could be improved on a compromise
solution in the medium term. 

As have been mentioned in~\ref{sec:Background}, several experimental
works have been carried out for fluoride, chloride an nitrate salts
(single, binary and ternary mixtures) since the ARE project in ORNL.
The working fluid selection must take into account different circumstances,
so there are different candidate salt mixtures for different applications.
Firstly, in case of primary nuclear coolant the ideal salt must have:
(i) a melting point well below the coldest point in the circuit, (ii)
a boiling point and thermal stability well above any credible accident
condition temperatures, (iii) a low vapor pressure, (iv) a low viscosity
at operating temperature, (v) a large heat capacity, good thermal
conductivity, low Prandtl number, and (vi) a large change in specific
volume with temperature to effectively drive natural circulation cooling.
Chemical and neutron behavior may be also constraints in the fluid
selection, so that the neutron-absorption cross section of coolant
should be lower enough to guarantee stability under radiation. Moreover,
the coolant selection have to ensure compatibility with structural
materials of core, loop, and piping system~\citep{Greene2011}.

Fuel-salts require in addition an adequate actinide solubility at
the working core temperature, and the outlined features have also
to be properly identified for the fuel mixture.

Secondary coolants have no nuclear constraints, and the evaluation
criteria is mainly based on heat transport performance. But the same
chemical system should be employed in both sides of HTX, so that compatibility
of structural alloy with two different species is not required at
the working temperature. Therefore, requirements are simpler than
for primary coolant: low melting and boiling point, low vapor pressure,
thermal stability and conductivity, low viscosity, large heat capacity,
material compability, and preferably non/low toxicity. Some of them
have been discussed as ``Figure of merits'' (FOM) by Idaho National
Laboratory (INL)~\citep{kim2011development}. Whatever function proposed
(primary, fuel or secondary coolant), commercial availability and
industrial processing cost must also be take in account. The cost
and especial care demanded for Be handling have been also referred
occasionally by experimental researchers as an additional use constraint,
at least during initial knowledge development of high temperatures
performance~\citep{williams2005chemical,williams2006assess69}.

As oxygen-containing salts, nitrite-nitrate salt mixtures are not
suitable as primary coolants~\citep{ingersoll2006trade}. Although
latest reports agree on avoid chlorides for MSR, and nitrites-nitrates
are also unlisted~\citep{Benes2012}, both of them are frequently
referred for heat exchange coolants in energy applications. Williams~\citep{williams2006assess69}
recommended the ternary eutectic $\mathrm{LiCl-KCl-MgCl_{2}}$ for
additional study because of its potential and raw-material cost, and
this salt has been recognized also by Beneš and Konings~\citep{Benes2012,Benes2009}
as FLiNaK alternative HTF in the VHTR concept. As mentioned above,
current Fluoride-Salt-Cooled High-Temperature (FHR) version includes
FluZirK as secondary coolant~\citep{holcomb2012current}.

This section have checked the most important criteria used for molten
salt selection, that can be briefly summarized in the following general
ideas:
\begin{itemize}
\item high volumetric heat capacity compared with other coolants
\item mandatory ${{}^7}$Li purity requirements in nuclear applications
\item the cost is lower for chlorides, and higher for fluorides, fluorobotaes
and Li or Be containing mixtures
\item Zr provides actinides solubility, oxygen getter capability and red-ox
control to salt mixtures, but also have undesirable activation products
\item O-containing salts are not suitable as primary coolants
\item Zr-fluorides are also being studied as secondary coolants, and chlorides
may be a low cost alternative to take into account.
\end{itemize}

\section{Molten salt thermo-physical properties}

In this section, a review of the most relevant thermo-physical properties
for design calculations is shown for the most common and recent coolant
or HTS composition choices:
\begin{enumerate}
\item FLUORIDES: $2\mathrm{LiF}$-$\mathrm{BeF_{2}}$, (hereafter FLiBe),
$\mathrm{LiF}$-$\mathrm{NaF}$-$\mathrm{KF}$ (hereafter FLiNaK),
$\mathrm{LiF}$-$\mathrm{NaF}$-$\mathrm{BeF_{2}}$ (hereafter FLiNaBe),
$\mathrm{NaF}$-$\mathrm{NaBF_{4}}$ (hereafter NaFNaB), $\mathrm{KF}$-$\mathrm{ZrF_{4}}$
(hereafter FluZirK)
\item CHLORIDES: $\mathrm{KCl}$-$\mathrm{MgCl_{2}}$ (hereafter CloKMag)
\item NITRATES: $\mathrm{NaNO_{3}}$-$\mathrm{KNO_{3}}$ (hereafter Solar
Salt), and $\mathrm{NaNO_{3}}$-$\mathrm{NaNO_{2}}$-$\mathrm{KNO_{3}}$
(hereafter Hitec$^{\lyxmathsym{\textregistered}}$ )
\end{enumerate}
Selected properties, by order, are summarized for each salt mixture:
melting point, density, viscosity, heat capacity, thermal conductivity.
Most usual molar compositions are shown in brackets, in~\ref{tab:Melting-point-proposed}.
When available, possible differences have been checked by using different
compositions of the same salt by the same work or author, in order
to analyze its behavior in molar terms and the differences among the
temperature range of measurement. 

Phase diagrams and useful properties have been also recently reviewed
by Beneš and Konings~\citep{Benes2009,Benes2012} for nuclear fission
applications, including critical discussion of recent values and measurements
of FLiBe, FLiNaK and NaFNaB. 

The currently useful mixtures for solar energy are mainly nitrates
and nitrites. Beneš et al.~\citep{Benes2010} used a encapsulation
technique with Solar Salt, measuring phase diagram. Ferri et al. proposed
selected properties of Solar Salt in RELAP5-3D code for solar parabolic
collectors~\citep{Ferri2008}, and Bauer et al. reviewed thermo-physical
correlations~\citep{Bauer_2012}. Both Solar Salt and Hitec$^{\lyxmathsym{\textregistered}}$
mixtures have been used in other heat transfer investigations~\citep{siegel2012thermal,schuller2012molten,codd2011direct,iverson2012thermal,mao2010heat}.

Other particular references are listed for each property, in order
to make a complete review of the cases. The general procedure has
involved a global plot of all the correlations found, verifying consistency
and agreement among them. The aim of this analysis is to create a
set of physical properties for technical use in system codes, computational
simulations, or experimental workbenches. Data may be also used to
observe dispersion and agreement in properties predictions, before
make decisions on critical parameters of design in energy projects.

\subsection{Melting point (\emph{M.P.},~\textmd{\textdegree{}}K)}

As has already been commented, phase diagrams for molten salts have
been continuously investigated and reviewed in other to detect the
most useful compositions for each mixture.~\ref{tab:Melting-point-proposed}
shows different values for melting point parameter for selected or
more common molar compositions. In the case of FLiNaBe there are not
many experimental values for the usually referred compositions (0.31-0.31-0.38)
or (0.33-0.33-0.33).

\noindent 
\begin{table*}
\caption{\label{tab:Melting-point-proposed}\protect \\
Melting point (\textdegree{}K) proposed or used by different authors
for all the studied salts.}

\noindent \centering{}%
\begin{tabular}{>{\raggedright}p{1cm}rcccccccc}
\toprule 
 & {\scriptsize Salt mixture~~} & {\tiny FLiBe} & {\tiny FLiNaK} & {\tiny FLiNaBe} & {\tiny NaFNaB} & {\tiny FluZirK} & {\tiny CloKMag} & {\tiny Solar Salt} & {\tiny Hitec$^{\lyxmathsym{\textregistered}}$}\tabularnewline
\cmidrule{2-10} 
\multirow{2}{1cm}{{\scriptsize Refer.}} & \multirow{2}{*}{{\scriptsize Usual Composition~~}} & {\tiny (0.66-} & {\tiny (0.465-0.115-} & {\tiny (0.31-0.31-} & {\tiny (0.08-} & {\tiny (0.58-} & {\tiny (0.68-} & {\tiny (0.66-} & {\tiny (0.07-0.49-}\tabularnewline
 &  & {\tiny -0.34)} & {\tiny -0.42)} & {\tiny -0.38)} & {\tiny -0.92)} & {\tiny -0.42)} & {\tiny -0.32)} & {\tiny -0.34)} & {\tiny -0.44)}\tabularnewline
\midrule 
\multicolumn{2}{l}{{\scriptsize Hoffman and Lones (1955)~\citep{hoffman1955fused}}} &  & {\scriptsize 727.4} &  &  &  &  &  & \tabularnewline
\midrule
\multicolumn{2}{l}{{\scriptsize Cohen and Jones (1957)~\citep{cohen1957viscosity}}} &  & {\scriptsize 727} &  &  &  &  &  & \tabularnewline
\midrule 
\multicolumn{2}{l}{{\scriptsize Grimes et al. (1958)~\citep{lane1958fluid}}} &  & {\scriptsize 727} & {\scriptsize 611} &  &  &  &  & \tabularnewline
\midrule 
\multicolumn{2}{l}{{\scriptsize Thoma (1959)~\citep{thoma1959phase}}} & {\scriptsize 727} & {\scriptsize 727} & {\scriptsize 588} &  & {\scriptsize 663} &  &  & \tabularnewline
\midrule 
\multicolumn{2}{l}{{\scriptsize Cantor et al. (1968)~\citep{Cantor_et_al1968}}} & {\scriptsize 731} &  &  & {\scriptsize 658} &  &  &  & \tabularnewline
\midrule 
\multicolumn{2}{l}{{\scriptsize Barton et al. (1971)~\citep{barton1971phase}}} &  &  &  & {\scriptsize 654} &  &  &  & \tabularnewline
\midrule 
\multicolumn{2}{l}{{\scriptsize Janz et al. (1972)~\citep{Janz1972}}} &  &  &  &  &  &  & {\scriptsize 495} & \tabularnewline
\midrule 
\multicolumn{2}{l}{{\scriptsize Cantor (1973)~\citep{Cantor1973}}} &  &  &  & {\scriptsize 658} &  &  &  & \tabularnewline
\midrule 
\multicolumn{2}{l}{{\scriptsize Janz et al. (1974)~\citep{Janz1974}}} & {\scriptsize 731.9} &  &  & {\scriptsize 657} &  &  &  & \tabularnewline
\midrule 
\multicolumn{2}{l}{{\scriptsize Janz et al. (1975)~\citep{Janz1975}}} &  &  &  &  &  & {\scriptsize 699} &  & \tabularnewline
\midrule 
\multicolumn{2}{l}{{\scriptsize Janz et al. (1978)~\citep{janz1978physical}}} & {\scriptsize 729} & {\scriptsize 727} &  & {\scriptsize 657} & {\scriptsize 693} & {\scriptsize 710} &  & \tabularnewline
\midrule 
\multicolumn{2}{l}{{\scriptsize Vriesema (1979)~\citep{vriesema1979aspects}}} &  & {\scriptsize 727} &  &  &  &  &  & \tabularnewline
\midrule 
\multicolumn{2}{l}{{\scriptsize Janz and Tomkins (1981)~\citep{janz1981physical}}} &  & {\scriptsize 727} &  &  &  & {\scriptsize 708} &  & {\scriptsize 415}\tabularnewline
\midrule 
\multicolumn{2}{l}{{\scriptsize Abe et al. (1981)~\citep{Abe1981}}} & {\scriptsize 732.1} &  &  &  &  &  &  & \tabularnewline
\midrule 
\multicolumn{2}{l}{{\scriptsize Janz and Tomkins (1983)~\citep{Janz1983}}} &  &  &  & {\scriptsize 657} &  &  &  & \tabularnewline
\midrule 
\multicolumn{2}{l}{{\scriptsize Mlynariková et al. (2008)~\citep{Mlynarikova2008}}} &  &  &  & {\scriptsize 658.7} &  &  &  & \tabularnewline
\midrule 
\multicolumn{2}{l}{{\scriptsize Beneš et al. (2010)~\citep{Benes2010}}} &  &  &  &  &  &  & {\scriptsize 496} & \tabularnewline
\bottomrule
\end{tabular}
\end{table*}

\subsection{Density ($\rho$,~kg/m\textthreesuperior{})}

\begin{figure*}
\centering{}\includegraphics[width=0.95\textwidth]{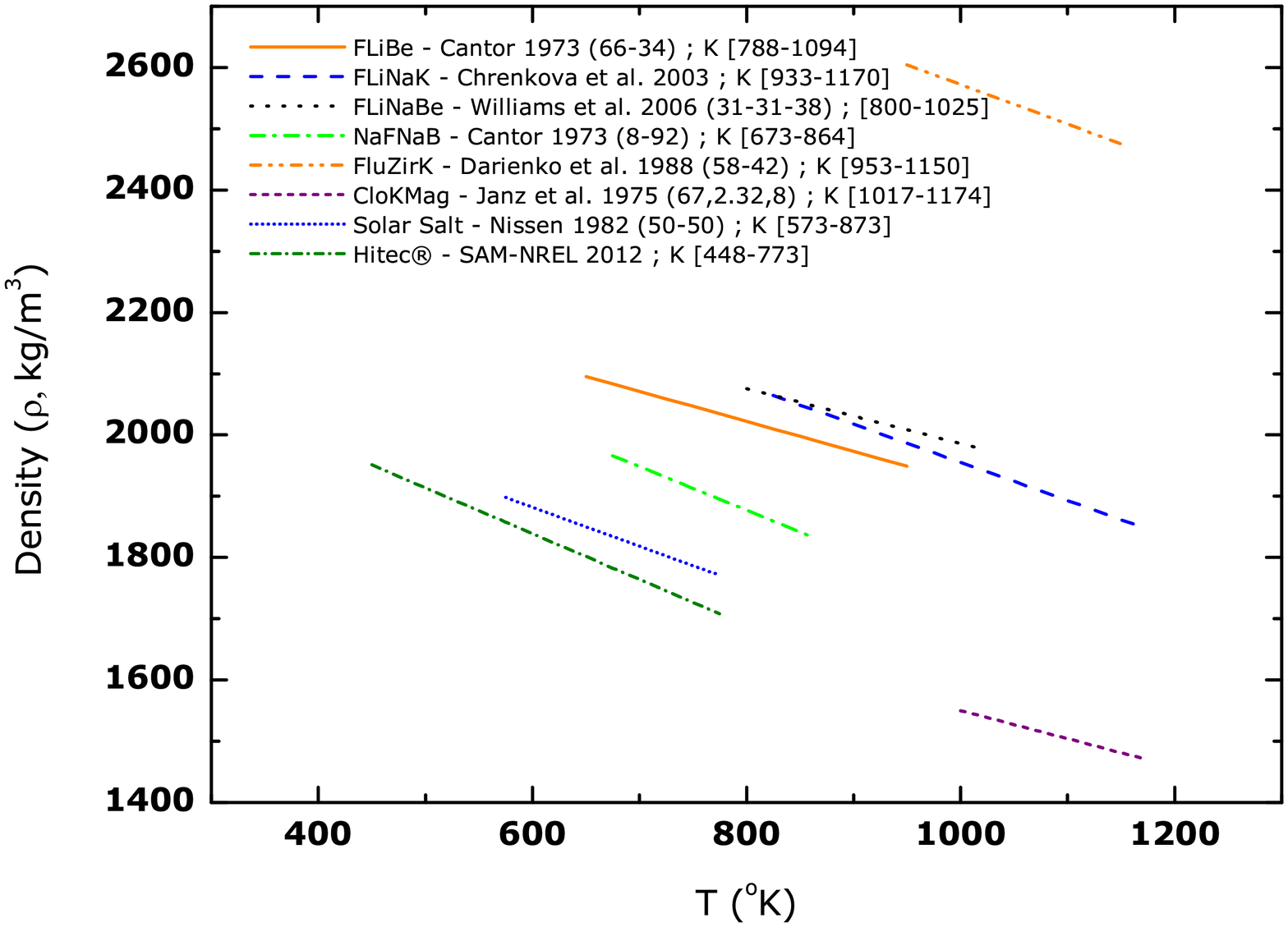}\caption{\label{fig:Density-compared}Graphical density comparison of the different
mixtures studied by selected correlations, showing a similar negative
slope.}
\end{figure*}

In general, a good agreement is verified by all selected mixtures
for density correlations. Recommended values are summarized in~\ref{tab:Density-functions},
and~\ref{fig:Density-compared}. Density of FLiBe has been studied
and experimented for different compositions since 1956. The slope
of the temperature function in the reviewed bibliography agrees with
very little variation. Changes in molar percentages do not suppose
great differences (e.g.,~\citet{powers1963physical,cantor_1969,Janz1974}),
with the exception of values offered by~\citet{lane1958fluid} for
50\%-50\% molar mixture, and recently cited by~\citet{Korkut2011}.
In this case, a significant disagreement is observed, and correlated
values are apparently too large, and the deviation from average is
over 11.8 \%.

\begin{figure*}
\noindent \begin{centering}
\subfloat[\label{fig:Density-functions-FLIBE}FLiBe.]{\noindent \begin{centering}
{\footnotesize \includegraphics[width=8.5cm]{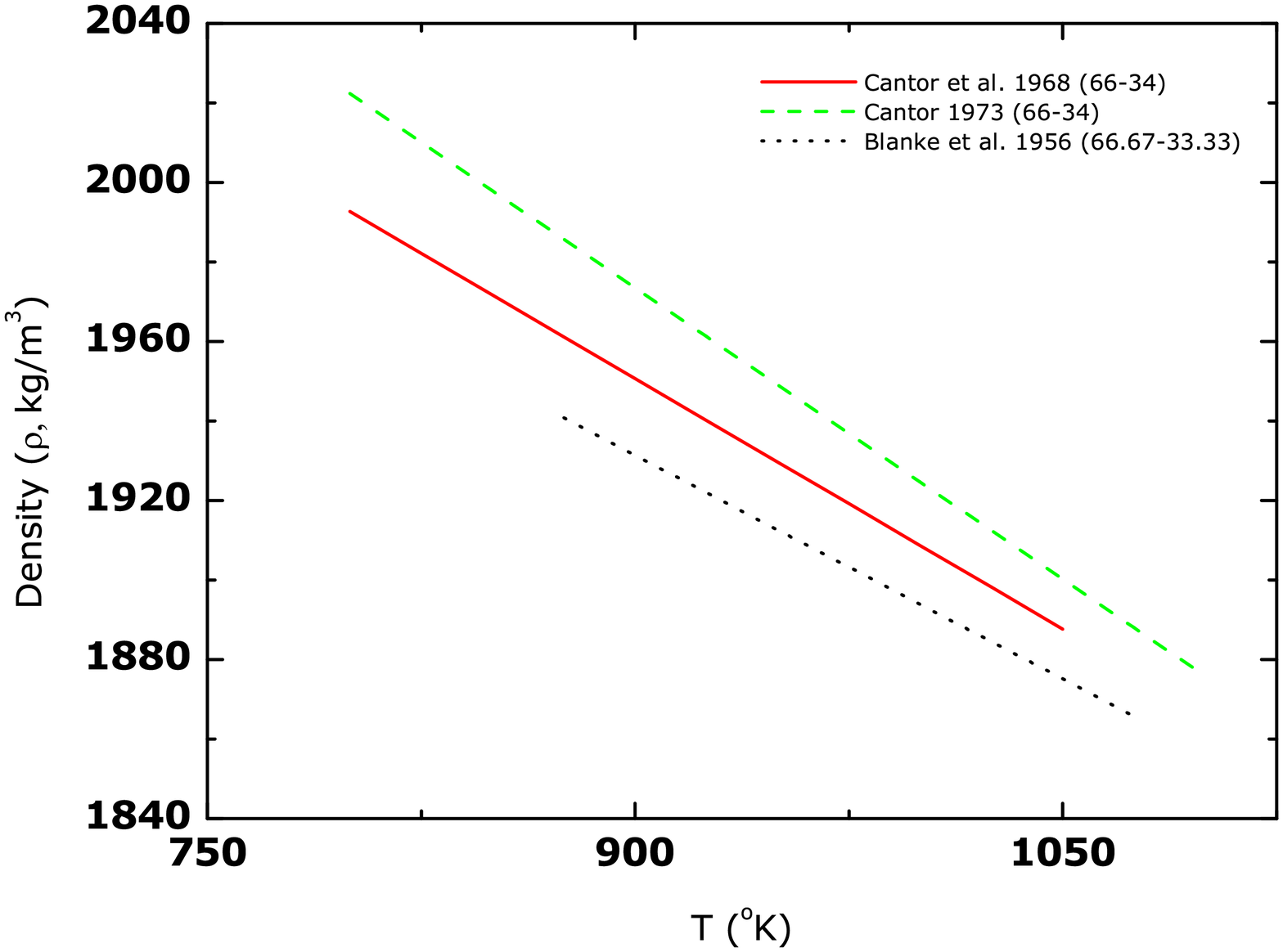}}
\par\end{centering}{\footnotesize \par}

\noindent \centering{}}\subfloat[\label{fig:Density-functions-FLINAK}FLiNaK.]{\noindent \begin{centering}
{\footnotesize \includegraphics[width=8.5cm]{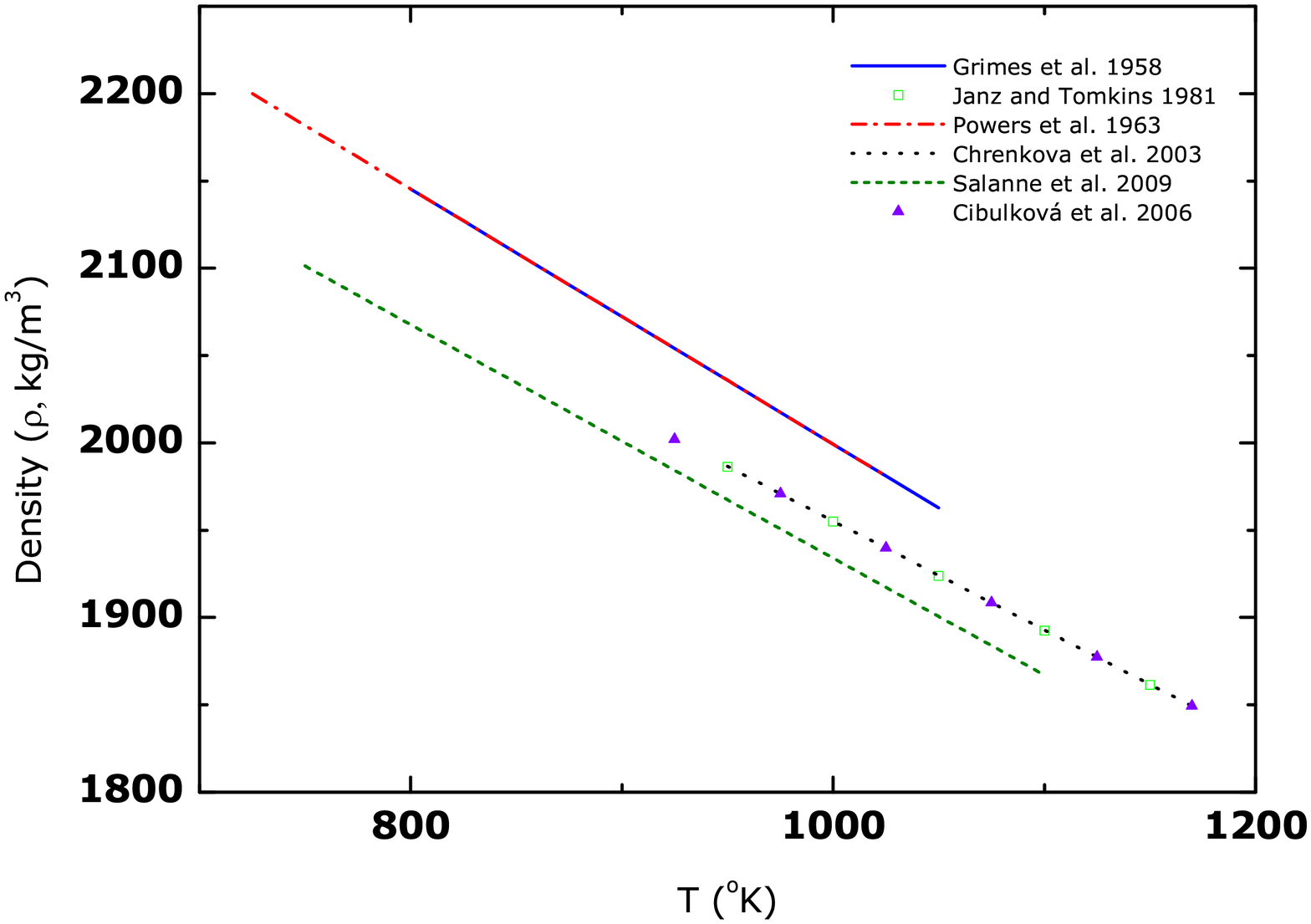}}
\par\end{centering}{\footnotesize \par}

\noindent \centering{}}
\par\end{centering}

\noindent \begin{centering}
{\footnotesize }\subfloat[\label{fig:Density-functions-FLINABE}FLiNaBe.]{\noindent \begin{centering}
{\footnotesize \includegraphics[width=8.5cm]{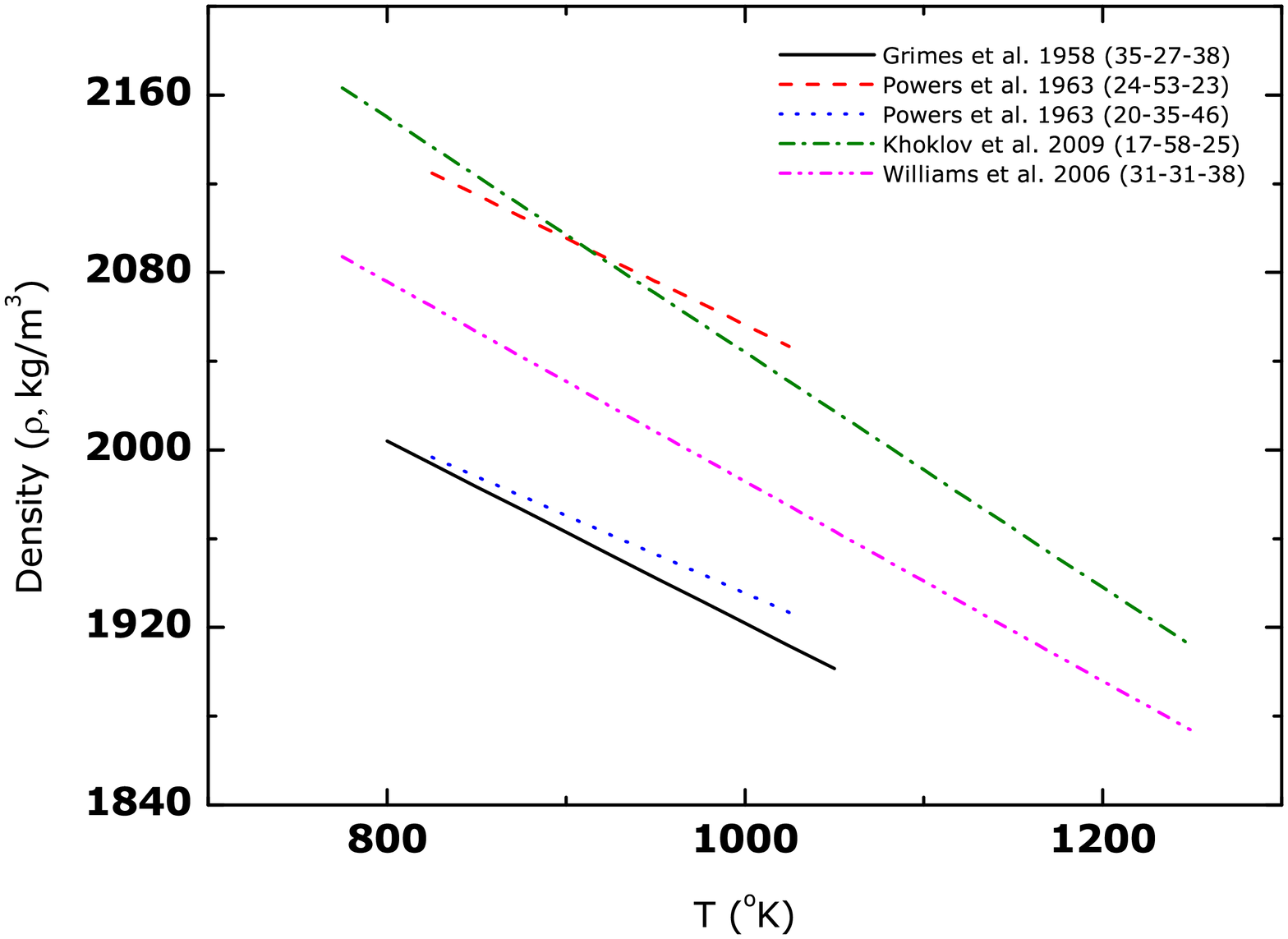}}
\par\end{centering}{\footnotesize \par}

}
\par\end{centering}{\footnotesize \par}

\caption{Comparison of density variation with respect to temperature for the
salts: FLiBe (a), FLiNaK (b), and FLiNaBe (c). Suggested values for
each three salts are very near to average, giving a deviation below
0.9 \%.}
\end{figure*}

The standard molar composition (0.66-0.34) has been correlated by~\citet{Blanke1956,Cantor_et_al1968}
and~\citep{Cantor1973}. \citet{Cantor1973} is recommended for use,
in agreement with \citet{Benes2012}. However, this last work does
not offer the correct temperature function, differing from the original
proposal. So best values are obtained for the range~$\mathrm{T\in[788-1094]}$
(see~\ref{eq:rho_flibe}). The deviation from the average is around
0.89 \% for the mentioned selected correlation, while the global standard
deviation for the three plotted values (\ref{fig:Density-functions-FLIBE})
is about 17.3.
\begin{equation}
\mathrm{\rho\,(kg/m\text{\textthreesuperior})=2413.03-0.4884\text{·}T(\text{\textdegree}\mathrm{K})}\label{eq:rho_flibe}
\end{equation}

\noindent 

The density of eutectic FLiNaK has been experimentally measured and
estimated in different occasions, as reported by~\citet{lane1958fluid,janz1981physical,Chrenkova2003,Cibulkova2006a}
and~\citet{Salanne2009_26} for several temperature ranges. \citet{williams2006assess12,ambrosek2009current},
and~\citet{Korkut2011} used Grimes et al. as a reference. However,
and as suggested by~\citet{Benes2009}, data of Chrenková et al.
are near and parallel to ideal density behavior (lightly minor than
ideal). Although Salanne et al. estimation is nearest in timeline,
it is based on Molecular Dynamics (MD) simulations and the calculated
values are more distant if compared with the ideal behavior. An excellent
correlation between the function reported by Chrenková and the one
from Janz and Tomkins can also be observed; the standard deviation
is 17.6 for the six plotted functions (\ref{fig:Density-functions-FLINAK}).
Results of Cibulková et al.~\prettyref{eq:rho_flinak} are close
to Chrenková et al. Both of them are suggested as best currently values,
giving a 0.38 \% deviation from average values. Any of this two functions
may be extrapolated for the whole range~$\mathrm{T\in[933-1170]}$:

\begin{equation}
\mathrm{\rho\,(kg/m\text{\textthreesuperior})=2579.3-0.6237\text{·}T(\text{\textdegree}\mathrm{K})}\label{eq:rho_flinak}
\end{equation}

The usually suggested composition for ternary FLiNaBe in recent papers
is (0.31-0.31-0.38) or (0.33-0.33-0.33). The nearest correlation found
was published by~\citet{williams2006assess12}, by the method of
additive molar volumes. Moreover,~\citet{Benes2012} suggested a
different composition for the fuel matrix in MSR using FLiNaBe$\mathrm{-AnF{}_{4}}$
(0.203-0.571-0.212-0.013). Other researchers have proposed other expressions,
e.g.~\citet{lane1958fluid} for (0.35-0.27-0.38),~\citet{powers1963physical}
for (0.20-0.35-0.46), and~\citet{Khokhlov2009} for (0.22-0.567-0.213)
among others. All of them offer a similar slope (\ref{fig:Density-functions-FLINABE}).
The differences among the compositions result in a standard deviation
value of 64.9, but estimations like those of Williams et al.~(\ref{eq:rho_flinabe})
are in halfway regarding the others (only 0.07 \% of deviation from
the average) and are suggested for the temperature range~$\mathrm{T\in[800-1025]}$:

\begin{equation}
\rho\,(\mathrm{kg/m\text{\textthreesuperior})=2435.85-0.45\text{·}T(\text{\textdegree}\mathrm{K})}\label{eq:rho_flinabe}
\end{equation}

\begin{figure*}
\noindent \begin{centering}
{\footnotesize }\subfloat[\label{fig:Density-functions-NAFNAB}NaFNaB. ]{\begin{centering}
{\footnotesize \includegraphics[width=8.5cm]{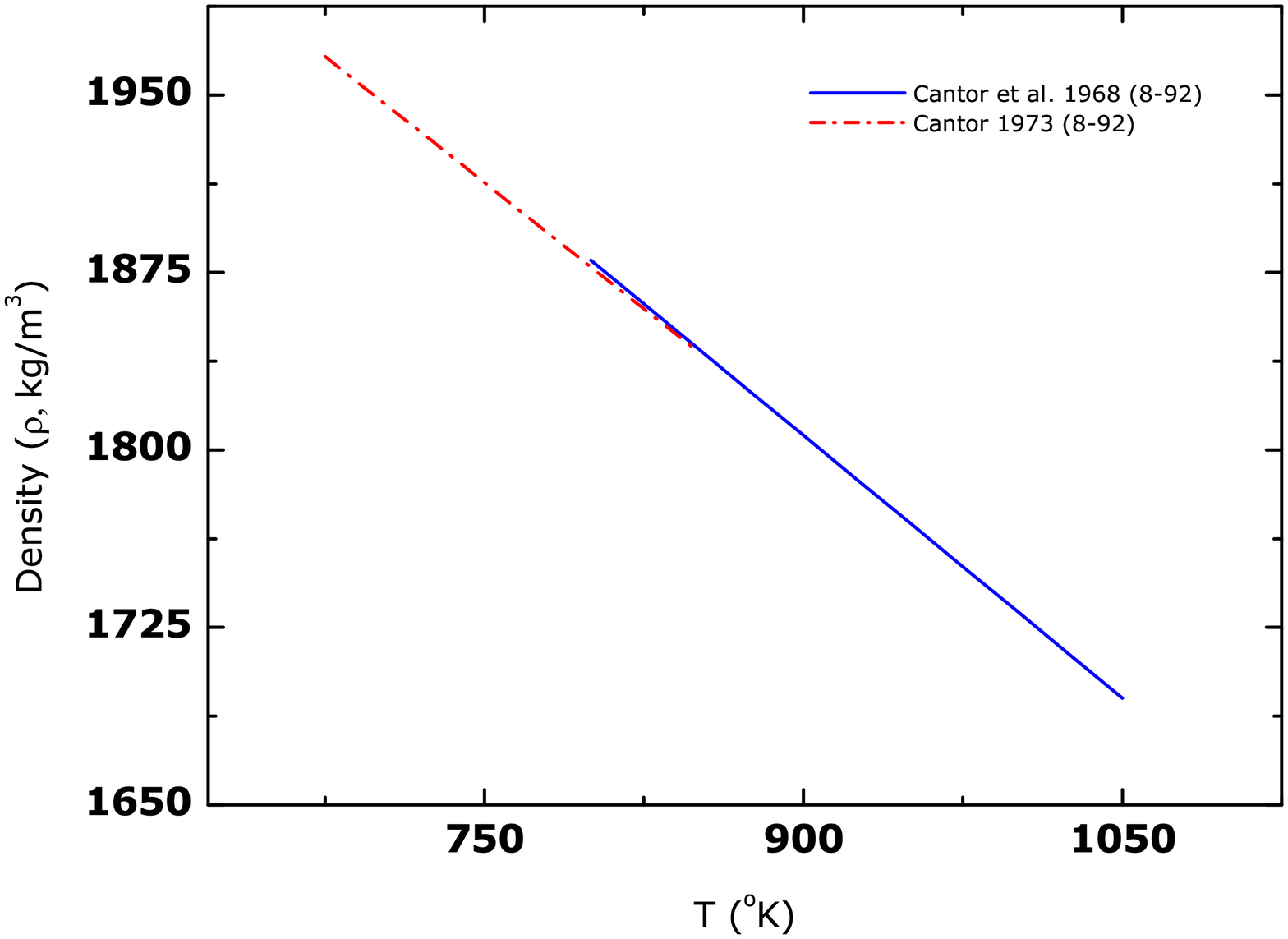}}
\par\end{centering}{\footnotesize \par}

\centering{}{\footnotesize }}\subfloat[\label{fig:Density-functions-FLUZIRK}FluZirK.]{\begin{centering}
{\footnotesize \includegraphics[width=8.5cm]{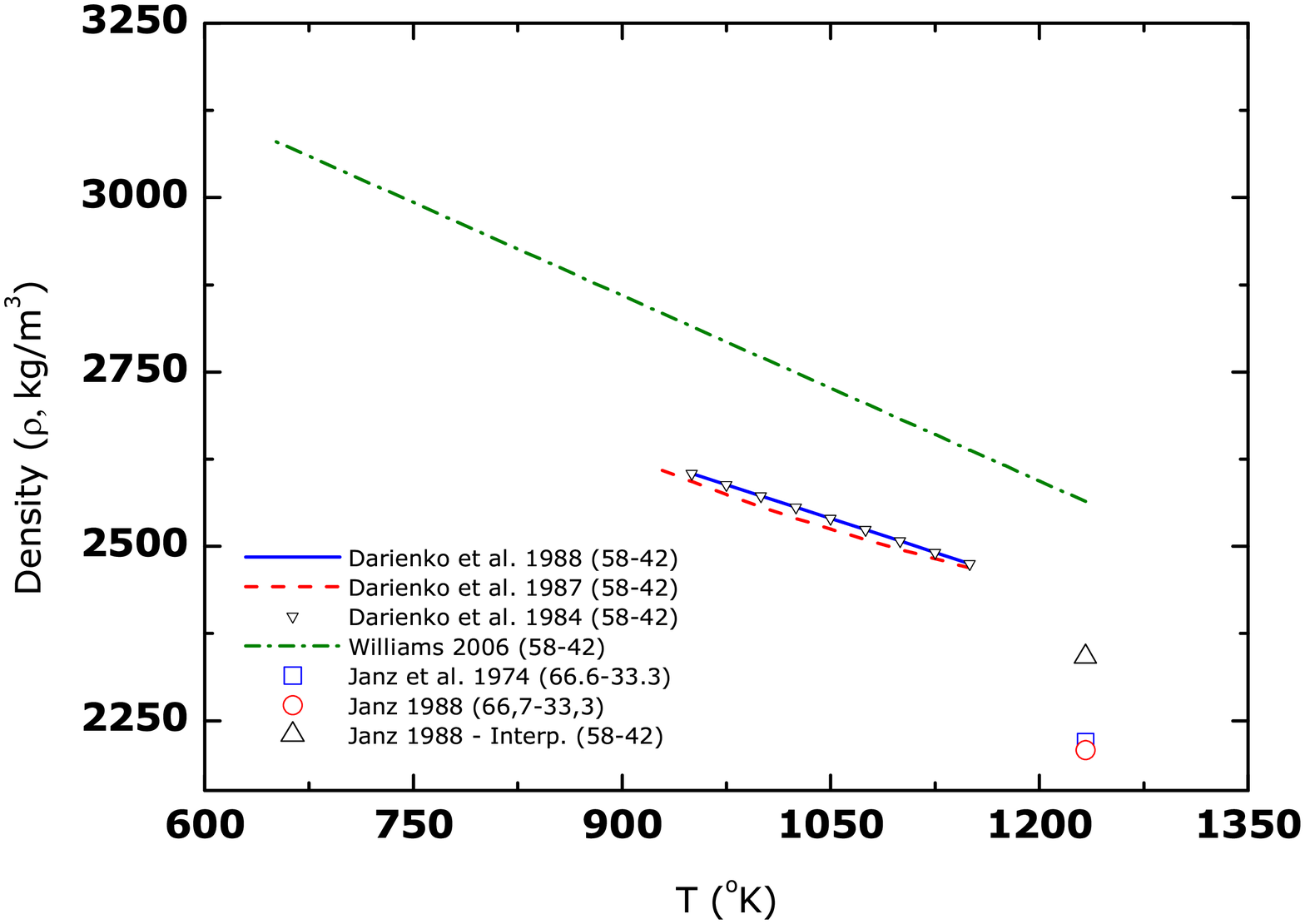}}
\par\end{centering}{\footnotesize \par}

\centering{}}
\par\end{centering}

\noindent \begin{centering}
{\footnotesize }\subfloat[\label{fig:Density-functions-CLOKMAG}CloKMag.]{\begin{centering}
{\footnotesize \includegraphics[width=8.5cm]{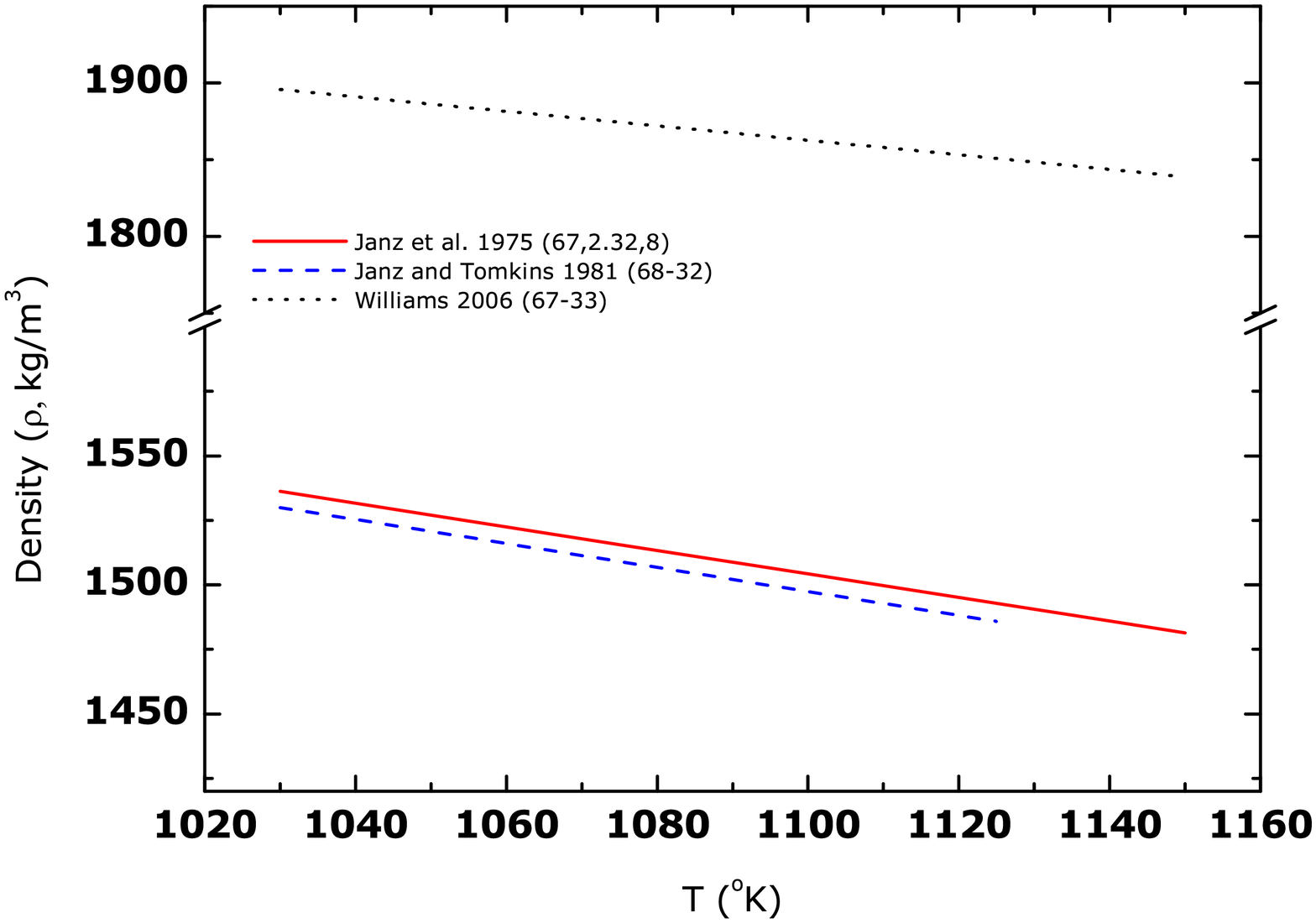}}
\par\end{centering}{\footnotesize \par}

\centering{}{\footnotesize }}
\par\end{centering}{\footnotesize \par}

\caption{Reviewed density correlations for NaFNaB (a), FluZirK (b) and CloKMag
(c). A very good agreement is observed for NaFNaB (a). Plot for FluZirK
(b) includes two different values correlated by Janz at 1233.2 \textdegree{}K
for the (0.67-0.33) molar ratio, and a extrapolated value for the
(0.58-0.42) composition. Williams expressions give too high density
values in the case of FluZirK and CloKMag mixtures, when compared
with other measurements.}
\end{figure*}

Two correlations have been found for NaFNaB,~\citet{Cantor_et_al1968}
and~\citep{Cantor1973}. Both equations are similar (\ref{fig:Density-functions-NAFNAB})
giving 1.47 for standard deviation at the overlap interval, and results
can be estimated by the second one for~$\mathrm{T\in[673-864]}$
in order to obtain differences from the average minor than 0.06 \%:

\begin{equation}
\rho\,\mathrm{\mathrm{(kg/m\text{\textthreesuperior})}=2446.2-0.711\text{·}T(\text{\textdegree}\mathrm{K})}\label{eq:rho_nafnab}
\end{equation}

\noindent 

FluZirK is being used recently in latest FHR concepts and proposals
with (0.58-0.42) molar relation. For this salt, no correlations were
made till 1988 to our best knowledge. Several data were formerly reported
by~\citet{Janz1974} and then by~\citet{Janz1988} at 1233.2 \textdegree{}K.
Different estimations were provided by~\citet{Darienko1984,Darienko1988}
from 33 to 80 mol\% of $\mathrm{ZrF_{4}}$, and also an estimation
may be calculated through the kinematic and dynamic viscosity relationship
given in~\citet{Darienko1987}. The graphical analysis of the parameters
given in the latter one shows an anomalous value for the 33 mol\%
$\mathrm{ZrF_{4}}$ compositions (whose standard deviation is higher,
as for the 60 and 80 mol\%). But, if we interpolate $\rho$ between
40 and 45 \% for the usual 42 mol\% the final curve fits with the
others, which shows the accuracy in this molar concentration range.
More recently a new correlation was published by~\citet{williams2006assess69}.
Studying the similarity among them (\ref{fig:Density-functions-FLUZIRK}),
the slope is almost the same in all functions. Darienko et al. results
are in a very good agreement in the three reported papers. However,
the equation given by Williams is quantitatively far from the others.
Indeed, standard deviation grows to 95.8 if taking into account Williams
correlation, while deviation goes to 7.34 when calculating this value
without this expression. When plotting all values together, proposals
of~\citet{Darienko1988}~\eqref{eq:rho_fluzirk} are more coherent
to those initially published by Janz et al. and lately by Janz. The
deviation from average is about 0.17 \% if we ignore Williams correlation;
but this value raises to 1.7 \% if we use this expression for average
estimation. Hence, although Williams expression has been used in recent
papers such as~\citet{anderson2012molten}, the most suitable correlation
for the range $\mathrm{T\in[953-1150]}$ reads as follows:

\begin{equation}
\rho\,\mathrm{\mathrm{(kg/m\text{\textthreesuperior})}=3217.44-0.6453\text{·}T(\text{\textdegree}\mathrm{K})}\label{eq:rho_fluzirk}
\end{equation}

Classical works of~\citet{Janz1975,janz1981physical} included several
expressions to estimate the density for CloKMag, from 25 to 42.2 mol\%
of $\mathrm{MgCl_{2}}$. Also~\citet{williams20066additional} has
reported another expression, but values obtained with it are too far
from the other ones (\ref{fig:Density-functions-CLOKMAG}). Standard
deviation is about 4.64 if avoiding Williams correlation, while this
value raises up to 221.6 when taking into account this expression.
Hence, Janz et al.~\eqref{eq:rho_clokmag} is suggested for future
works in the range $\mathrm{T\in[1017-1174]}$, with a 0.22 \% deviation
from average (without Williams):

\begin{equation}
\rho\,\mathrm{\mathrm{(kg/m\text{\textthreesuperior})}=2007-0.4571\text{·}T(\text{\textdegree}\mathrm{K})}\label{eq:rho_clokmag}
\end{equation}

\begin{figure*}
\begin{centering}
\subfloat[\label{fig:Density-functions-SOLAR_SALT}Solar Salt.]{\begin{centering}
{\footnotesize \includegraphics[width=8.5cm]{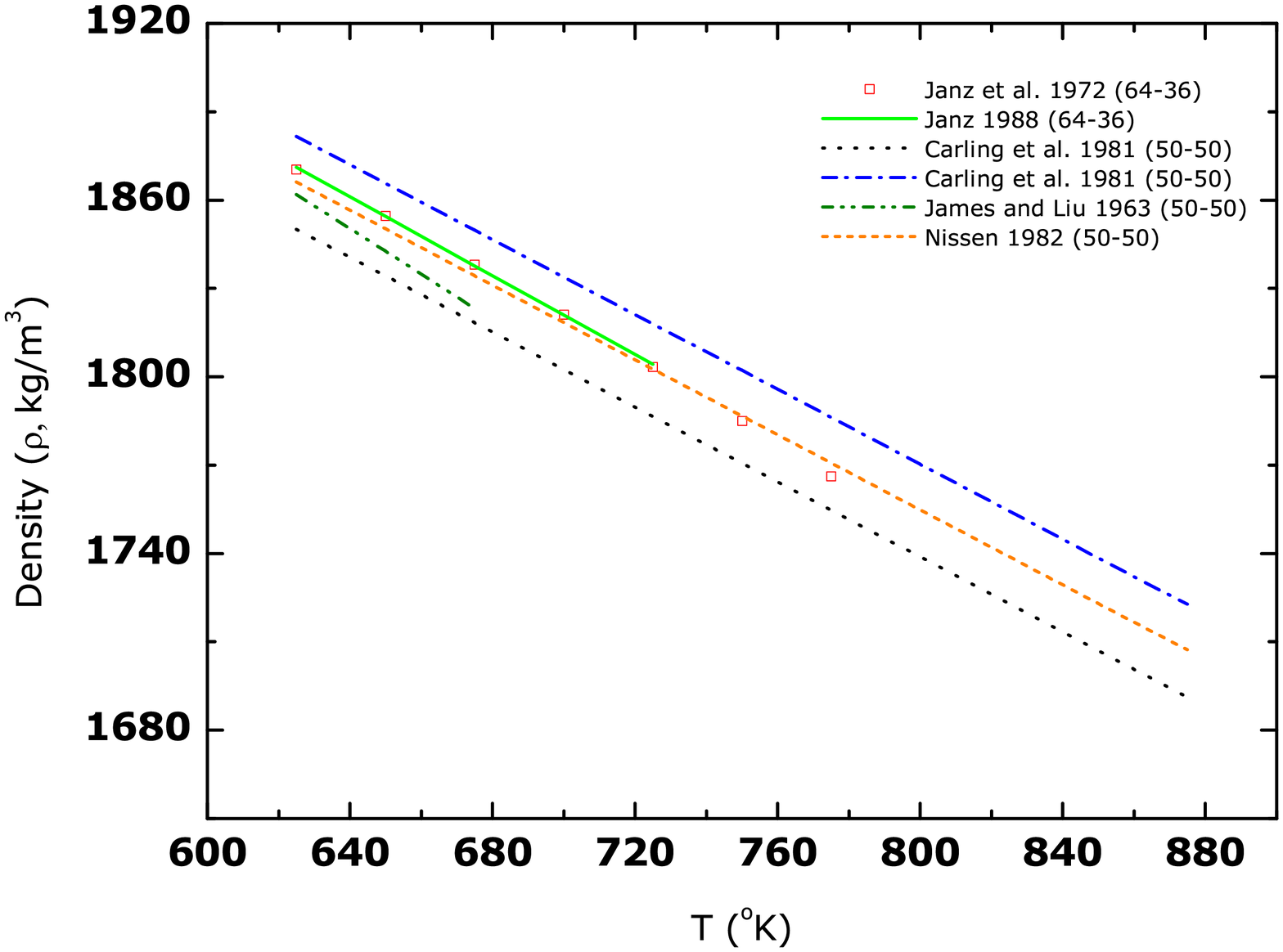}}
\par\end{centering}{\footnotesize \par}

}\\
\subfloat[\label{fig:Density-functions-HITEC}Hitec$^{\lyxmathsym{\textregistered}}$
.]{\begin{centering}
\includegraphics[width=8.5cm]{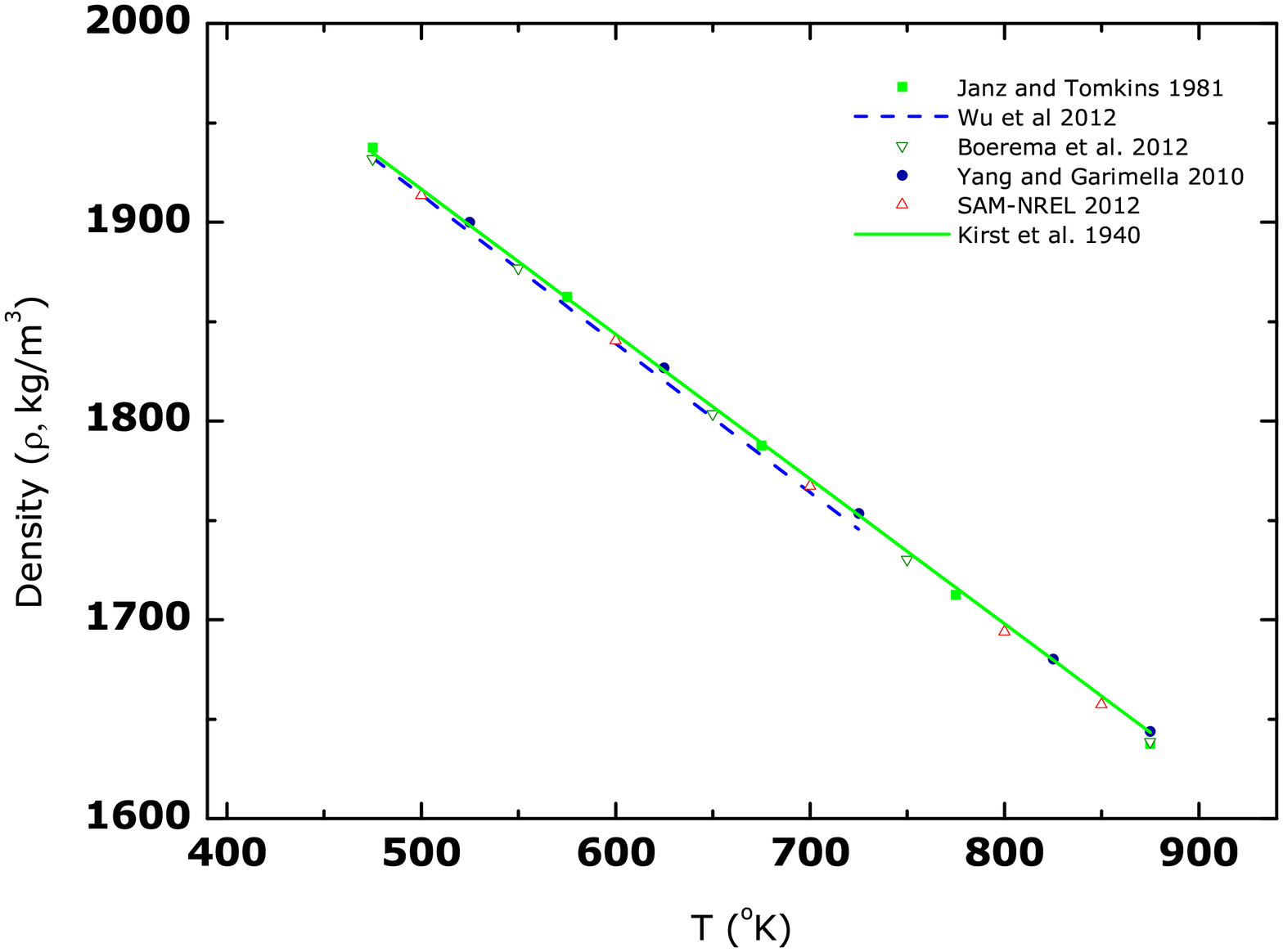}
\par\end{centering}

}
\par\end{centering}

\caption{Density correlations for nitrates: Solar Salt (a), and Hitec$^{\lyxmathsym{\textregistered}}$
mixture (b).}
\end{figure*}

Parameters for the so called Solar Salt (also known as draw salt)
have been either published for the equimolar composition (near 1:1
molar ratio), and for the cheaper commercial mixture (0.64-0.36),
or 60-40 wt\%. However, there are no significant differences in terms
of density. This useful mixture was reported by~\citet{Janz1972}
giving a second order equation in terms of temperature, and which
allows to take into account the molar percentage of $\mathrm{KNO_{3}}$.
\citet{Janz1988} includes linear expressions for the density from
30 to 50 mol\% of $\mathrm{KNO_{3}}$. The eutectic composition of
this salt was also studied by~\citet{James1963,Carling1981} and~\citet{Nissen1982}.
Carling et al. reported two correlations, taking into account the
thermal decomposition of nitrate to nitrite at extended time experiments.
This maybe the reason of the second order proposal of Janz et al.,
but calculated values do not exactly agree in both cases. The correlation
given by Nissen has been chosen by~\citet{Zavoico2001} at Sandia
National Laboratories, and is the current reference for the System
Advisor Model (SAM) at the National Renewable Energy Laboratory (NREL).
James and Liu values are slightly below from all the others, and the
curve has also a different slope. For the two possibilities given
by Carling et al., Nissen appears as the average value (\ref{fig:Density-functions-SOLAR_SALT})
with 0.03 \% of deviation, and 13.10 as global standard deviation.
As the values are so closer for the (0.64-0.36) commercial salt, negligible
error is done by using the same correlation for both mixtures. The
higher range of measurement reported by Nissen~\eqref{eq:rho_solar_nissen}
respect to Janz et al. suggests best accuracy. Therefore, for $\mathrm{T\in[573-873]}$
and the (0.50-0.50) composition, the Nissen expression is as follows:

\begin{equation}
\rho\,\mathrm{\mathrm{\mathrm{(kg/m\text{\textthreesuperior})}}=2263.641-0.636\text{·}T(\text{\textdegree}\mathrm{K})}\label{eq:rho_solar_nissen}
\end{equation}

The last salt reviewed is the commercially denoted as Hitec$^{\lyxmathsym{\textregistered}}$,
which has been largely used as HTF by chemical industry. Several studies
can be mentioned, with very little differences (\ref{fig:Density-functions-HITEC})
among their mathematical expression and nearly the same after calculating
values, e.g.,~\citet{Kirst1940} (as cited by~\citet{Gaune1982}),~\citet{janz1981physical,Yang2010,Wu2012}
and~\citet{Boerema2012}. Additionally, values of SAM for the Hitec$^{\lyxmathsym{\textregistered}}$
composition~\citep{NREL} have been correlated~\eqref{eq:rho_hitec}.
This last system is being commonly used as standard, giving a 0.09
\% deviation from the average. Using all the mentioned expressions,
a value of 2.55 is calculated as standard deviation. Hence the recommended
correlation for the temperature range $\mathrm{T\in[448-773]}$ reads:

\begin{equation}
\rho\,\mathrm{\mathrm{\mathrm{(kg/m\text{\textthreesuperior})}}=2279.799-0.7324\text{·}T(\text{\textdegree}\mathrm{K})}\label{eq:rho_hitec}
\end{equation}

\noindent 

Summarizing for density property,~\eqref{tab:Density-functions}
shows the selected functions for each salt with the calculated percentage
of deviation from the average values. Best fit is obtained for nitrates,
followed by NaFNaB and FLiNaBe salts. Nevertheless, tabulated expressions
provide acceptable estimations, even for FLiBe and FLiNaK (which show
the higher scattering among reviewed data). A graphical sketch of
selected functions is plotted in~\ref{fig:Density-compared}. This
figure provides a quickly view for all the studied salts: FluZirK
appears as the heaviest one, while CloKMag is the lightest of them.
Molten salt density decreases as temperature increases in all the
cases analyzed.

\noindent 
\begin{table*}[p]
\caption{\label{tab:Density-functions}\protect \\
Density correlations $\mathrm{(kg/m^{3})}$ as temperature function
suggested for studied salts, including references and deviation from
the global average data.}

\noindent \centering{}%
\begin{tabular}{cccccc}
\toprule 
{\scriptsize Salt mixture} & {\scriptsize Reference} & {\scriptsize Ref. num.} & {\scriptsize Selected correlation} & {\scriptsize Temp. Range} & {\scriptsize \% Dev.}\tabularnewline
\midrule
\midrule 
{\scriptsize FLiBe} & {\scriptsize Cantor 1973 (0.66-0.34)} & {\scriptsize ~\citep{Cantor1973}} & {\scriptsize 2413.03-0.4884·T} & {\scriptsize {[}788-1094{]}} & {\scriptsize 0.89 \%}\tabularnewline
\midrule 
{\scriptsize FLiNaK} & {\scriptsize Chrenkova et al. 2003 (0.465-0.115-0.42)} & {\scriptsize ~\citep{Chrenkova2003}} & {\scriptsize 2579.3-0.624·T} & {\scriptsize {[}933-1170{]}} & {\scriptsize 0.38 \%}\tabularnewline
\midrule 
{\scriptsize FLiNaBe} & {\scriptsize Williams et al. 2006 (31-31-38)} & {\scriptsize ~\citep{williams2006assess12}} & {\scriptsize 2435.85-0.45·T} & {\scriptsize {[}800-1025{]}} & {\scriptsize 0.07 \%}\tabularnewline
\midrule 
{\scriptsize NaFNaB} & {\scriptsize Cantor 1973 (0.08-0.92)} & {\scriptsize ~\citep{Cantor1973}} & {\scriptsize 2446.2-0.711·T} & {\scriptsize {[}673-864{]}} & {\scriptsize 0.06 \%}\tabularnewline
\midrule 
{\scriptsize FluZirK} & {\scriptsize Darienko et al. 1988 (0.58-0.42)} & {\scriptsize ~\citep{Darienko1988}} & {\scriptsize 3217.44-0.6453·T} & {\scriptsize {[}953-1150{]}} & {\scriptsize 0.17 \%}\tabularnewline
\midrule 
{\scriptsize CloKMag} & {\scriptsize Janz et al. 1975 (0.672-0.328)} & {\scriptsize ~\citep{Janz1975}} & {\scriptsize 2007-0.4571·T} & {\scriptsize {[}1017-1174{]}} & {\scriptsize 0.22 \%}\tabularnewline
\midrule 
{\scriptsize Solar Salt} & {\scriptsize Nissen 1982 (0.50-0.50)} & {\scriptsize ~\citep{Nissen1982}} & {\scriptsize 2263.628-0.636·T} & {\scriptsize {[}573-873{]}} & {\scriptsize 0.03 \%}\tabularnewline
\midrule 
{\scriptsize Hitec$^{\lyxmathsym{\textregistered}}$} & {\scriptsize SAM-NREL 2012 (0.07-0.49-0.44)} & {\scriptsize ~\citep{NREL}} & {\scriptsize 2279.799-0.7324·T} & {\scriptsize {[}448-773{]}} & {\scriptsize 0.09 \%}\tabularnewline
\bottomrule
\end{tabular}
\end{table*}

\noindent 

\subsection{Dynamic viscosity ($\eta$,~Pa·s)}

Most of selected salts follow an Arrhenius behavior in their temperature
range. However, some nitrate salts are susceptible to decompose at
high temperatures or extended time heat expositions. The non-Arrhenius
behavior has been reported for some chloride mixtures by~\citet{Boon1989},
discussed by~\citet{Nissen1982} for equimolar Solar Salt, and lately
described by~\citet{Bradshaw2010} for some multicomponent nitrates.
The Fulcher expression (\ref{eq:vfth}, also known as VFTH) can be
used for correlation in this cases, but several authors still use
a simple polynomial regression.

\begin{equation}
\mathrm{\log_{10}\,(\eta)=A+\frac{B}{T-T_{0}}}\label{eq:vfth}
\end{equation}
\citet{cantor_1969}~also conjectured about the non-Arrhenius behavior
in FLiBe when applying low temperatures, which can be extrapolated
for other molten salts. However, working range is higher to preclude
this, far enough from the double of temperature of ideal glass transition
point.

\begin{figure*}
\begin{centering}
\subfloat[\label{fig:Viscosity-FLIBE}FLiBe.]{\begin{centering}
\includegraphics[width=8.5cm]{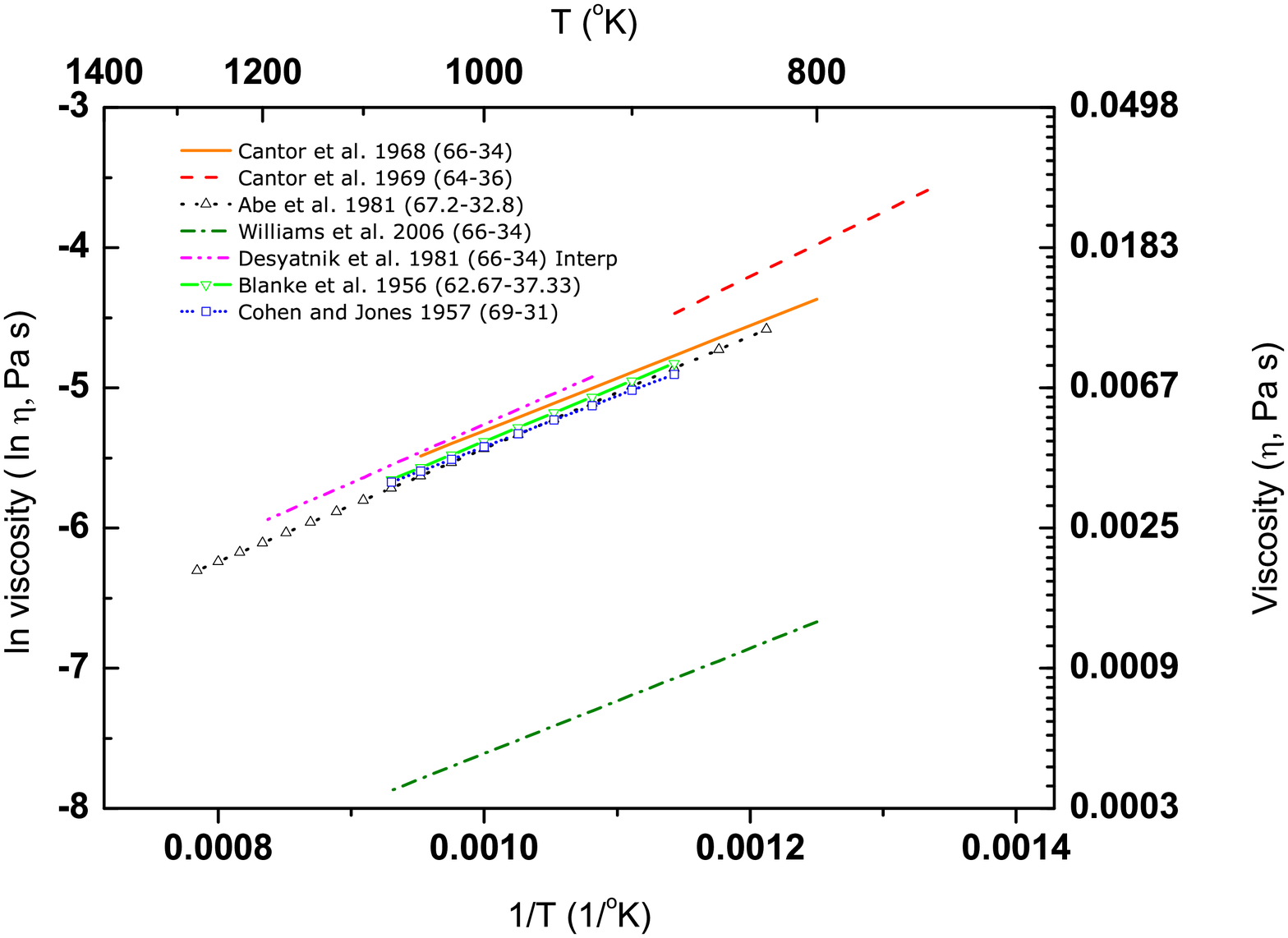}
\par\end{centering}

\begin{centering}

\par\end{centering}

}\subfloat[\label{fig:Viscosity-FLINAK}FLiNaK.]{\begin{centering}
\includegraphics[width=8.5cm]{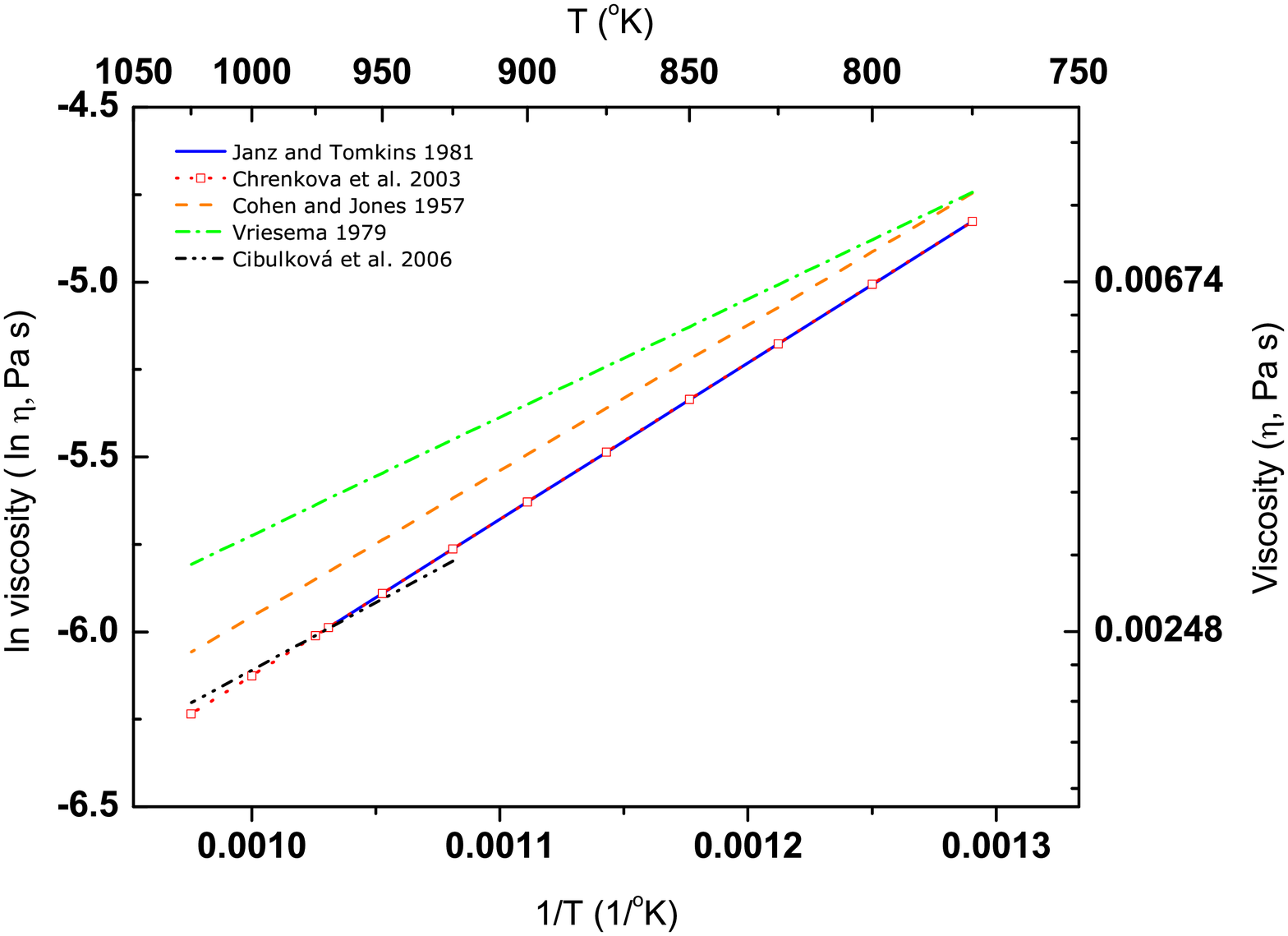}
\par\end{centering}

\begin{centering}

\par\end{centering}

}
\par\end{centering}

\centering{}\subfloat[\label{fig:Viscosity-FLINABE}FLiNaBe.]{\begin{centering}
\includegraphics[width=8.5cm]{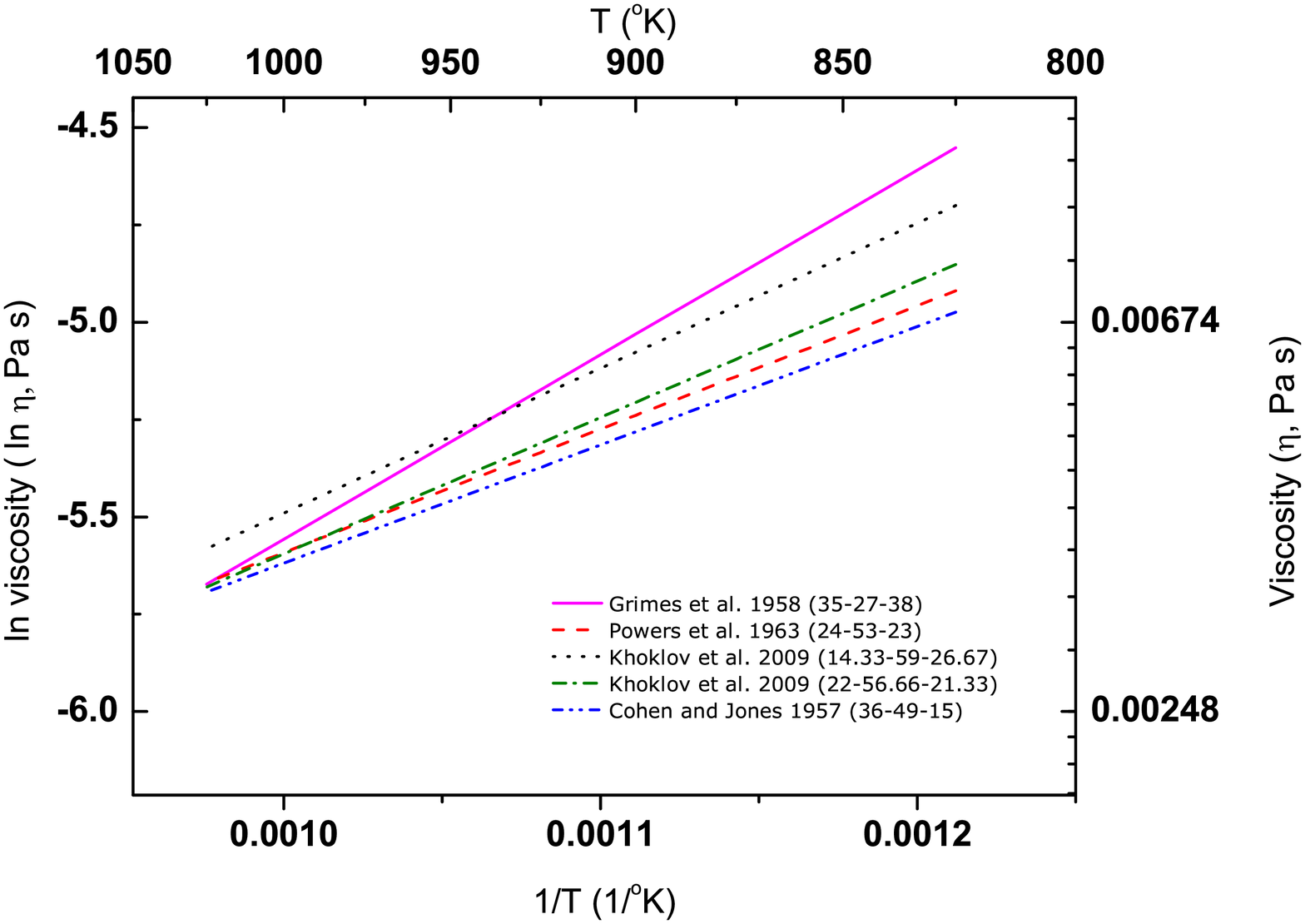}
\par\end{centering}

}\caption{Comparison of viscosity variation with respect to temperature for
FLiBe (a), FLiNaK (b), and FLiNaBe (c), according to empirical correlations
of different studies and authors.}
\end{figure*}

Around FLiBe viscosity, Salanne\emph{ }et al\emph{.}~\citep{salanne2007conductivity}
explained the reason of pure~$\mathrm{BeF_{2}}$ high viscosity based
in a MD study. As~$\mathrm{BeF_{2}}$ concentration increases in
a FLiBe solution, different species are created, resulting in a polymer
of several $\textrm{BeF}_{4}^{2-}$ units at the highest percentages.
Different investigations have been carried out about this physical
property. \citet{Blanke1956,Cantor_et_al1968,desyatnik1981viscosity}
measured viscosity for several molar concentrations of $\mathrm{BeF_{2}}$,
while~\citet{cohen1957viscosity,cantor_1969} and~\citet{Abe1981}
studied only one mixture. \citet{williams2006assess12} gave a correlation
based on \citet{Cantor_et_al1968}, but the expression is one order
of lower magnitude (\ref{fig:Viscosity-FLIBE}). The global standard
deviation grows from 0.0004 to 0.0034 when including Williams expression.
\citet{Janz1974} and \citet{janz1981physical} already used Cantor
et al. correlations for the molten salt database.

Globally, the agreement is nearly perfect~\citep{Benes2009} among
all values with the exception of Williams correlation, showing a rising
curve when plotted all data in a X (\% $\mathrm{BeF_{2}}$), Y (log
$\eta$) graph. For $\mathrm{T>1050}$, although viscosity becomes
almost constant, data are near parallel (\ref{fig:Viscosity-FLIBE}).
Percentage of deviation from the average is around 4.97 \% for Cantor
et al. and $\mathrm{T\in[800-1050]}$, and near to 7 \% for Abe at
al. at the same temperature interval. From these results, \citet{Abe1981}
is suggested for $\mathrm{T>1050}$~\textdegree{}K, while \citet{Cantor_et_al1968}~\eqref{eq:nu_flibe}
is selected for the range $\mathrm{T\in[800-1050]}$:

\begin{equation}
\eta\,\mathrm{(Pa\cdot s)=0.000116\text{·}exp(\frac{3755}{T(\text{\textdegree}\mathrm{K})})}\label{eq:nu_flibe}
\end{equation}

\citet{cohen1957viscosity} offered measurements for FLiNaK viscosity,
later reported by~\citet{lane1958fluid,powers1963physical,williams20066additional,Korkut2011}.
\citet{vriesema1979aspects} used a different data (kinematic viscosity
from a private communication of Oye, H. A.), and these values have
been correlated using the same density of his heat transfer experiments.
More recently, results provided by~\citet{Chrenkova2003} and correlation
listed by~\citet{janz1981physical} give nearly the same numbers
(\ref{fig:Viscosity-FLINAK}), and both of them can be over-ranged
by the equation of~\citet{Cibulkova2006a} up to 1163 \textdegree{}K.
After a comparison, and taking into account the discussed argument
for FLiNaK density, we suggest the correlation of Cherenková et al.~\eqref{eq:nu_flinak1}
for the range $\mathrm{T\in[773-1163]}$ (assuming negligible extrapolation
error because of nearness to Cibulková et al. values). However, if
using all reviewed data, the deviation from the average raises to
a 9.87 \% due to the closeness of all values:

\begin{equation}
\eta\,\mathrm{(Pa\cdot s)=0.0000249\text{·}10^{(\frac{1944}{T(\text{\textdegree}\mathrm{K})})}}\label{eq:nu_flinak1}
\end{equation}

The viscosity of FLiNaBe was measured by~\citet{cohen1957viscosity}
for different compositions, then reported by~\citet{lane1958fluid}
for (0.35-0.27-0.38) molar mixture, and also correlations were given
by~\citet{powers1963physical} for some other compositions. \citet{Khokhlov2009}
made calculations using additive law of molar volumes of simple $\mathrm{LiF}$
and $\mathrm{BeF_{2}}$, with binary $\mathrm{LiF-BeF_{2}}$ and $\mathrm{NaF-BeF_{2}}$.
As discussed by~\citet{Zherebtsov2001} the experimental data of
early ORNL measurements are in good agreement with the more recent
values. In addition,~\citet{Ignatiev2002} plotted a multi-comparison
graph showing a very good agreement with the Institute of High Temperature
Electrochemistry (IHTE) modeling equations used in the International
Science \& Technology Center (ISTC) \#1606-Project, when temperatures
are over 600\textdegree{}K. MSR application has been reviewed by \citet{Benes2012}
pointing to a recommended molar composition (0.22-0.5666-0.2133),
but they reproduced the equation corresponding to (0.1433-0.59-0.2667)
which must be taken into account. In other nuclear applications, the
most referred promising mixture is near to the equimolar. To the author's
knowledge, no investigation has been performed on this particular
salt.

All reviewed correlations show a good agreement for temperatures higher
to 950\textdegree{}K (\ref{fig:Viscosity-FLINABE}), where the composition
dependence do not imply so much scattering. Therefore, searching for
a global expression, accuracy of predictions for the range $\mathrm{T\in[823-1023]}$
is legitimated with the following approach, which give a 12.35 \%
of deviation from the overall average: 

\begin{equation}
\eta\,\mathrm{(Pa\cdot s)=0.0000338\text{·}exp(\frac{4738}{T(\text{\textdegree}\mathrm{K})})}\label{eq:nu_flinabe}
\end{equation}

\begin{figure*}
\begin{centering}
\subfloat[\label{fig:Visco-NaFNaB}NaFNaB.]{\begin{centering}
\includegraphics[width=8.5cm]{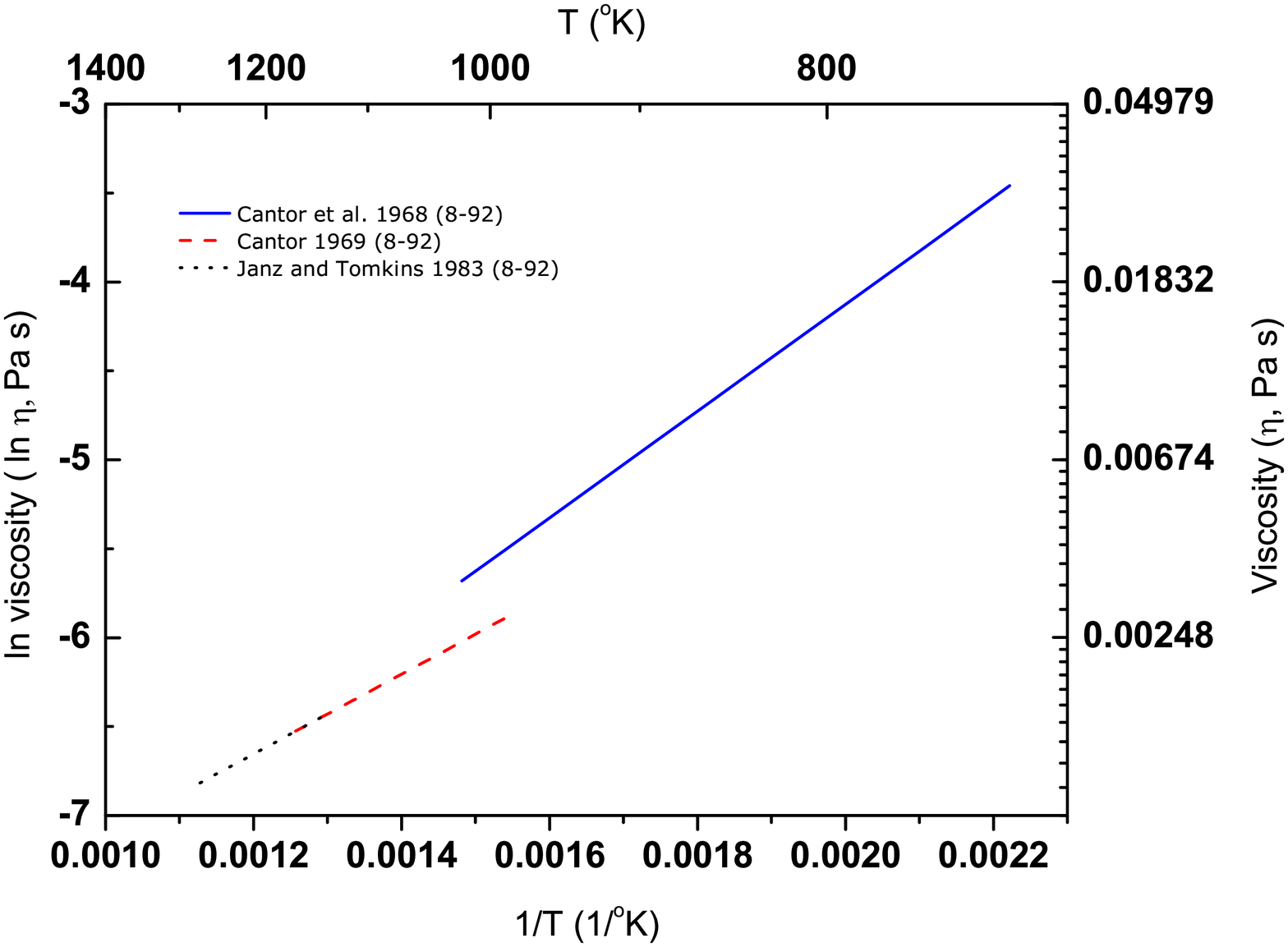}
\par\end{centering}

\begin{centering}

\par\end{centering}

}\subfloat[\label{fig:Visco-FluZirK.}FluZirK.]{\begin{centering}
\includegraphics[width=8.5cm]{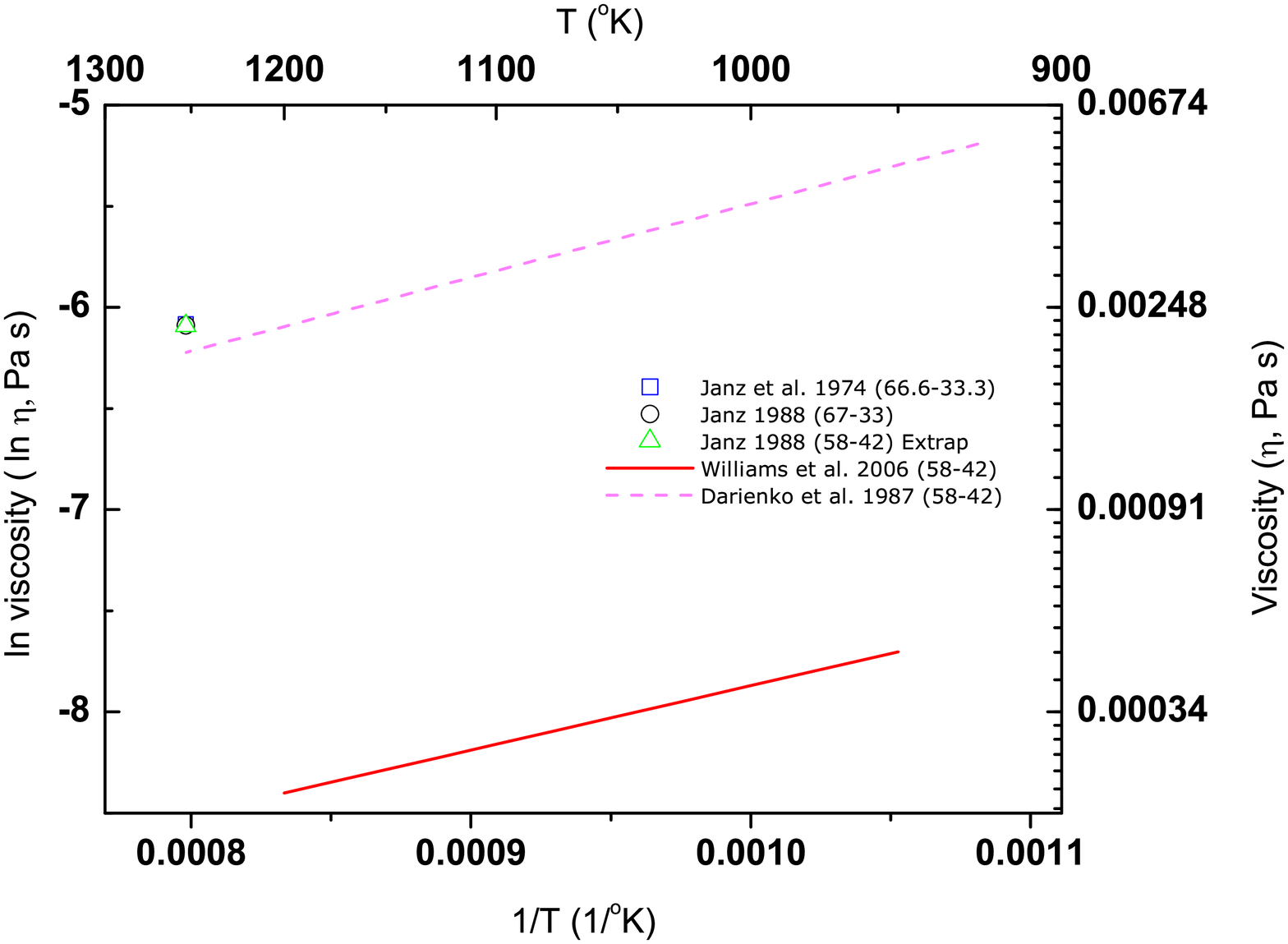}
\par\end{centering}

\begin{centering}

\par\end{centering}

}
\par\end{centering}

\centering{}\subfloat[\label{fig:Visco-CloKMag.}CloKMag.]{\centering{}\includegraphics[width=8.5cm]{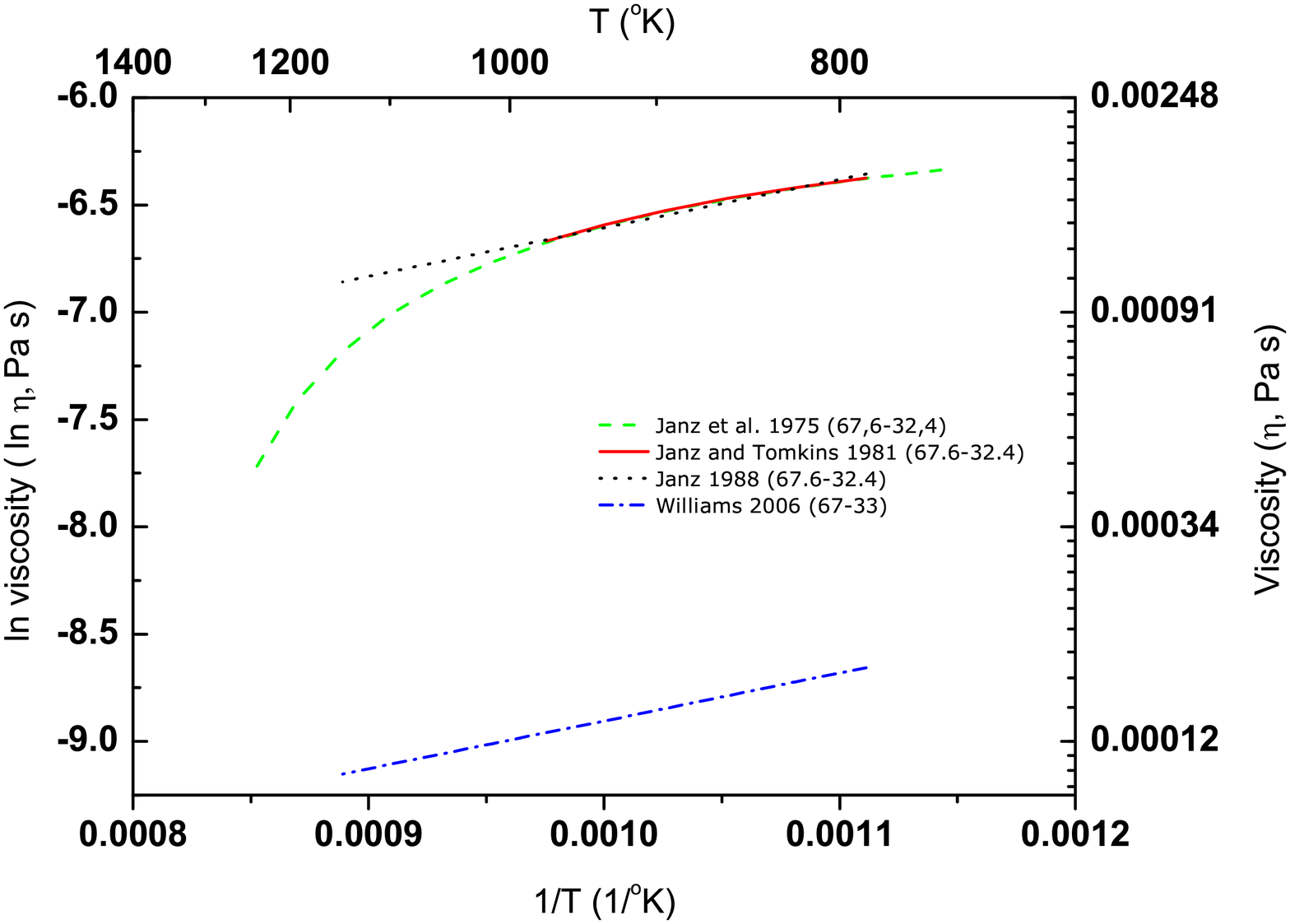}}\caption{Agreement among the reviewed correlations for NaFNaB viscosity (a).
Graphical comparison (b) among different proposed functions and values
correlated by Janz at 1253 \textdegree{}K for FluZirK viscosity. Functions
of temperature for CloKMag viscosity (c). The last two (b \& c) include
the anomalous values obtained with the expressions reported by Williams.}
\end{figure*}

Two formulations have been reported for NaFNaB at different temperature
ranges. \citet{Cantor_et_al1968} made a extrapolation for the behavior
of sodium iodide based in experimental values for NaFNaB published
by~\citet{Wittenberg1968} at Mound Laboratory. This latter correlation
was defined for $\mathrm{T\in[400-700]}$. Thereafter~\citet{Cantor1969}
gave a new equation for the range$\mathrm{T\in[682-810]}$, which
has been subsequently cited in later reports as~\citet{Cantor1973,Janz1974,Janz1988}.
\citet{Janz1983} expanded the temperature window between 799-906
\textdegree{}K, but reviewing this reference there are no apparently
reasons to include this new temperature range (\ref{fig:Visco-NaFNaB}).
So we suggest the function reported by~\citet{Cantor1969} but only
for the range $\mathrm{T\in[682-810]}$. Reviewed data give 0.003
for standard deviation at the overlap interval, while suggested expression
shows a 8.97 \% of deviation from average in the same interval:

\begin{equation}
\eta\,\mathrm{(Pa\cdot s)=0.0000877\text{·}exp(\frac{2240}{T(\text{\textdegree}\mathrm{K})})}\label{eq:nu_nafnab}
\end{equation}

\noindent 
\citet{Janz1974}~proposed viscosity values for two compositions
of FluZirK at 1253\textdegree{}K. Later,~\citet{Janz1988} reported
an expression to calculate this property at the same temperature for
a wide range of compositions, between 0-33,3 mol\% of $\mathrm{ZrF_{4}}$.
More recently,~\citet{Darienko1987} measured FluZirK viscosity for
concentrations from 0 to 80 mol\% of $\mathrm{ZrF_{4}}$, and~\citet{williams2006assess12}
proposed a exponential correlation for the promising (0.58-0.42) composition
as a function of temperature. Differences have been analyzed, after
interpolating data from Darienko et al. for the same (0.58-0.42) mixture~\eqref{eq:nu_fluzirk}.
Expression reported by Williams et al. shows noticeably lower values
the others, which means a 0.0022 global standard deviation and about
82.6 \% of deviance from the calculated average. However, our interpolation
points to very similar values as previously listed by Janz and Janz
et al. (\ref{fig:Visco-FluZirK.}). Hence, although Williams correlation
has been widely used and cited in many recent reports and papers (e.g.,~\citet{anderson2012molten,Kubikova2012,ALISIA_D-50,williams2006assess69,Samuel2009,Scheele2010,Sabharwall2011,sabharwall2011process}),
the following interpolated expression based in values of Darienko
et al. is suggested in present work for the range $\mathrm{T\in[921-1185]}$,
reducing deviation from the average to 45.24 \% when data of Williams
are still included in calculations:

\begin{equation}
\eta\,\mathrm{(Pa\cdot s)=0.0001084\text{·}10^{(\frac{1581.2}{T(\text{\textdegree}\mathrm{K})})}}\label{eq:nu_fluzirk}
\end{equation}

The viscosity of binary chloride CloKMag was correlated for different
molar concentrations by~\citet{Janz1975} and~\citet{janz1981physical},
giving one third order polynomial expression (including the usually
accepted 0.68-0.32 composition) and other standard Arrhenius formulation.
Also~\citet{Janz1988} and~\citet{williams2006assess69} reported
two different correlations for the mentioned (0.68-0.32) composition.
By calculating values we have compared all the possibilities (\ref{fig:Visco-CloKMag.}).
Williams expression appears with a lower order of magnitude, giving
a global 6.2·$10^{-4}$ standard deviation, while all the others are
in good agreement. Standard deviation is reduced to 5·$10^{-5}$ when
ignoring Williams correlation. Therefore,~\citet{Janz1988}~\eqref{eq:nu_clokmag}
is the recommended correlation for the whole range $\mathrm{T\in[900-1030]}$,
which means a 3.95 \% of deviation from the average:

\begin{equation}
\eta\,\mathrm{(Pa\cdot s)=0.0001408\text{·}exp(\frac{2261.3}{T(\text{\textdegree}\mathrm{K})})}\label{eq:nu_clokmag}
\end{equation}

\begin{figure}
\noindent \begin{centering}
\subfloat[\label{fig:Viscosity-SolarSalt}Solar Salt.]{\centering{}\includegraphics[width=8.5cm]{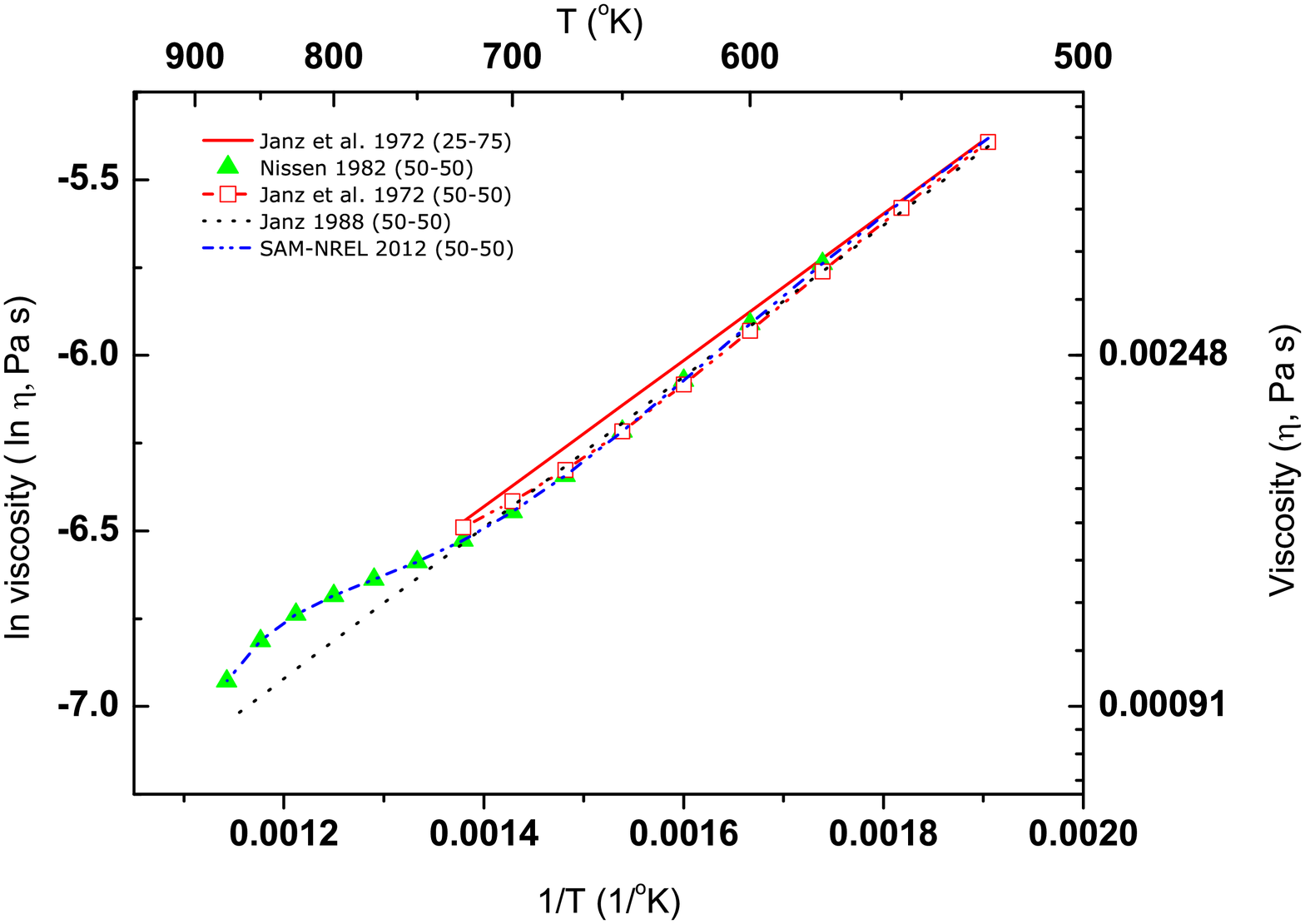}}\\
\subfloat[\label{fig:Viscosity-HITEC}Hitec$^{\lyxmathsym{\textregistered}}$.]{\centering{}\includegraphics[width=8.5cm]{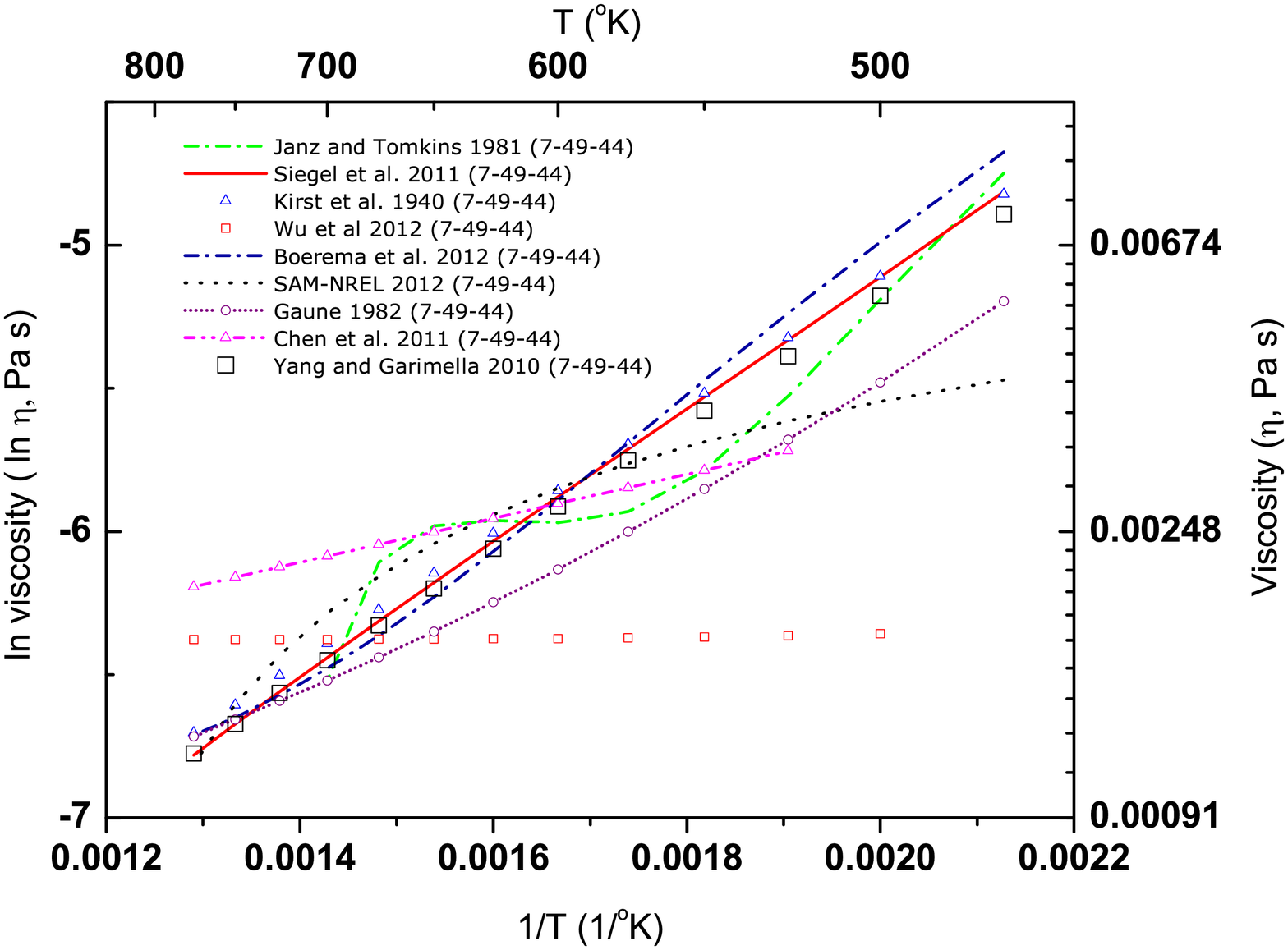}}
\par\end{centering}

\centering{}\caption{Graphical comparison of viscosity functions for Solar Salt (a) and
Hitec (b).}
\end{figure}

For the binary Solar Salt, negligible differences were found for viscosity
between equimolar and commercial compositions. Initial measurements
made by Murgulescu and Zuca were reported by~\citet{Janz1972} for
the range $\mathrm{T\in[525-725]}$ after a critical review. These
correlations included the 0, 25, 50, 75 and 100 mol\% of $\mathrm{NaNO{}_{3}}$,
with Arrhenius form for the two first and a third order polynomial
for the others. New Arrhenius expressions were reported by~\citet{Janz1988}
for the same molar composition cases, revising again the measurements
mentioned above. For the equimolar salt, experimental data was given
by~\citet{Nissen1982}, making a polynomial correlation used lately
by~\citet{Zavoico2001} for the Basis Document of the \emph{Solar
Power Tower}. Data offered in SAM~\citet{NREL} have been also correlated,
using polynomial~(\ref{eq:nu_solarsalt}) and Arrhenius forms. After
a comparison, data chart of SAM give the same values as Nissen for
a shorter temperature range (\ref{fig:Viscosity-SolarSalt}). A value
of 6E-05 has been calculated as global standard deviation. Apparently,
the equation given by Nissen shows the high temperature behavior of
nitrates regarding to decomposition (as mentioned previously), while
this chemical mechanism is not reflected in the Arrhenius form. In
any case, the effect of molar composition is almost negligible. The
following polynomial correlation is suggested for the whole range
$\mathrm{T\in[573-873]}$, using data of Nissen, and giving a 2.33
\% of deviation from the average values:

\begin{gather}
\eta\,\mathrm{\mathrm{(Pa\cdot s)=}0.07543937-2.77\cdot10^{-4}\cdot T(\text{\textdegree}\mathrm{K})}\label{eq:nu_solarsalt}\\
\mathrm{\mathrm{+3.49\cdot10^{-7}\cdot T^{2}(\text{\textdegree}\mathrm{K})-1.47\cdot10^{-10}\cdot T^{3}(\text{\textdegree}\mathrm{K})}}\nonumber 
\end{gather}

\noindent 

\noindent Different correlations have been published for the Hitec$^{\lyxmathsym{\textregistered}}$
mixture. \citet{Kirst1940} proposed a exponential behavior for temperatures
between 473-773 \textdegree{}K, while~\citet{Gaune1982} used a second
order function of temperature for the same range, with Arrhenius global
form. \citet{janz1981physical} reported a third order polynomial
correlation, but it shows a excessive slope at high temperatures regarding
the others (probably due to thermal decomposition of nitrates). More
recently, other authors have used other expressions, e.g.~\citet{Yang2010,Siegel2011,Chen2011,Boerema2012}
and~\citet{Wu2012}. Values of SAM~\citet{NREL} have been also
correlated to a linear expression in present work~\eqref{eq:nu_hitec}.
After a comparison, Siegel et al. gives approximately the same values
as Kirst et al. (\ref{fig:Viscosity-HITEC}).

Boerema et al. gives the higher viscosity of all the correlations
at low temperatures, as well as Wu et al. gives the lowest one, and
shows a crescent function of temperature. The expression provided
by Yang and Garimella is selected as the most representative. This
correlation was obtained by correlating Coastal Chemical data, giving
values in coherence with Solar Salt (also a nitrate mixture). Standard
deviation is calculated, giving a global 3·$10^{-4}$ value. Hence,
the suggested expresion for the range $\mathrm{T\in[525-773]},$ with
3.65 \% of deviation from the average, is as follows:

\begin{gather}
\eta\,\mathrm{\mathrm{(Pa\cdot s)=}\exp(-4.343-2.0143\text{·}}\label{eq:nu_hitec}\\
\mathrm{\mathrm{\text{·((}\ln(T(\text{\textdegree}\mathrm{K})-273)-5.011))}}\nonumber 
\end{gather}

\begin{figure*}
\centering{}\includegraphics[width=0.95\linewidth]{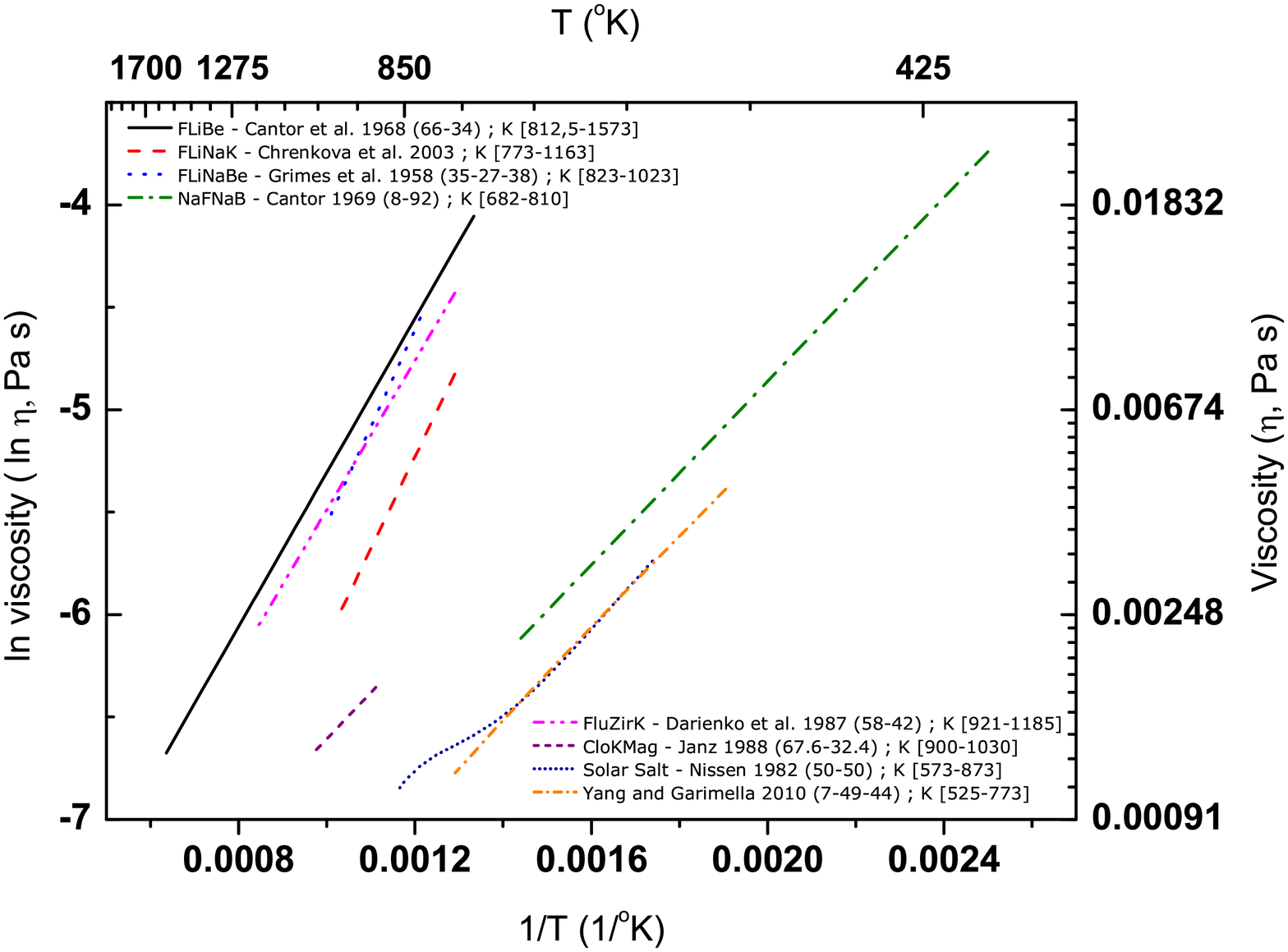}\caption{\label{fig:Viscosity-compared}Global plot of selected expressions
of viscosity for the different mixtures studied. A descent slope is
showed by all the salts. The lowest viscosity values, in studied temperature
intervals, correspond to nitrates and chlorides.}
\end{figure*}

In short, for dynamic viscosity, suggested expressions are listed
in~\ref{tab:Visco-functions} for all the studied mixtures. Data
for Solar salt give the smallest deviation from average values, while
correlations for NaFNab show the higher scattering by far (when compared
with the other salts). A graphical comparison is plotted in~\ref{fig:Viscosity-compared};
all functions show a descent slope. Data for FLiBe are valid in a
large temperature range, while viscosity correlations for the remainder
mixtures are only useful for a short interval.

\noindent 
\begin{table*}
\caption{\label{tab:Visco-functions}\protect \\
Viscosity correlations~(Pa·s) as temperature function, including
final deviation from the average values after ignoring both too high
as too low values.}

\noindent \centering{}%
\begin{tabular}{cccccc}
\toprule 
{\scriptsize Salt mixture} & {\scriptsize Reference} & {\scriptsize Ref. Num.} & {\scriptsize Selected correlation} & {\scriptsize Temp. Range} & {\scriptsize \% Dev.}\tabularnewline
\midrule
\midrule 
{\scriptsize FLiBe} & {\scriptsize Cantor et al. 1968 (0.66-0.34)} & {\scriptsize ~\citep{Cantor_et_al1968}} & {\scriptsize 1.16·$10^{-4}$·exp(3755/T) } & {\scriptsize {[}800-1050{]}} & {\scriptsize 4.97 \%}\tabularnewline
\midrule 
{\scriptsize FLiNaK} & {\scriptsize Chrenkova et al. 2003 (0.465-0.115-0.42)} & {\scriptsize ~\citep{Chrenkova2003}} & {\scriptsize 2.49·$10^{-5}$·10\textasciicircum{}(1944/T)} & {\scriptsize {[}773-1163{]}} & {\scriptsize 9.87 \%}\tabularnewline
\midrule 
{\scriptsize FLiNaBe} & {\scriptsize Grimes et al. 1958 (0.35-0.27-0.38)} & {\scriptsize ~\citep{lane1958fluid}} & {\scriptsize 3.38·$10^{-5}$·exp(4738/T)} & {\scriptsize {[}823-1023{]}} & {\scriptsize 12.35 \%}\tabularnewline
\midrule 
{\scriptsize NaFNaB} & {\scriptsize Cantor 1969 (0.08-0.92)} & {\scriptsize ~\citep{Cantor1969}} & {\scriptsize 8.77·$10^{-5}$·exp(2240/T)} & {\scriptsize {[}682-810{]}} & {\scriptsize 8.97 \%}\tabularnewline
\midrule 
{\scriptsize FluZirK} & {\scriptsize Darienko et al. 1987 (0.58-0.42)} & {\scriptsize ~\citep{Darienko1987}} & {\scriptsize 1.084·$10^{-4}$·10\textasciicircum{}(1581.2/T)} & {\scriptsize {[}921-1185{]}} & {\scriptsize 45.24 \%}\tabularnewline
\midrule 
{\scriptsize CloKMag} & {\scriptsize Janz 1988 (0.676-0.324)} & {\scriptsize ~\citep{Janz1988}} & {\scriptsize 1.408·$10^{-4}$·exp(2261.3/T)} & {\scriptsize {[}900-1030{]}} & {\scriptsize 3.95 \%}\tabularnewline
\midrule 
\multirow{3}{*}{{\scriptsize Solar Salt}} & \multirow{3}{*}{{\scriptsize Nissen 1982 (0.50-0.50)}} & \multirow{3}{*}{{\scriptsize ~\citep{Nissen1982}}} & {\scriptsize 0.075439-2.77·$10^{-4}$·(T-273)} & \multirow{3}{*}{{\scriptsize {[}573-873{]}}} & \multirow{3}{*}{{\scriptsize 2.33 \%}}\tabularnewline
 &  &  & {\scriptsize +3.49·$10^{-7}$·(T-273)\texttwosuperior{}} &  & \tabularnewline
 &  &  & {\scriptsize -1.474·$10^{-10}$·(T-273)\textthreesuperior{}} &  & \tabularnewline
\midrule 
{\scriptsize Hitec$^{\lyxmathsym{\textregistered}}$} & {\scriptsize Yang and Garimella 2010 (0.07-0.49-0.44)} & {\scriptsize ~\citep{Yang2010}} & {\tiny exp(-4.343-2.0143·(ln(T-273)-5.011))} & {\scriptsize {[}525-773{]}} & {\scriptsize 3.65 \%}\tabularnewline
\bottomrule
\end{tabular}
\end{table*}

\noindent 

\subsection{Thermal conductivity ($\lambda$, W · $\mathrm{m}^{-1}$ ·\textmd{
\textdegree{}}$\mathrm{K^{-1}}$)}

The measurement of thermal properties in molten salts has been pointed
out by many authors as an arduous task, specially in the case of conductivity.
At high temperatures, there are important uncertainties associated
to the heat transfer mode. \citet{NietodeCastro2004} also identified
the combined effect of factors involved in this kind of measurement,
analyzing the different methods available: (i) the sample purity and
homogeneity, (ii) thermal stability of salt, (iii) interaction between
the sample and both the surrounding gaseous atmosphere and the container
material, (iv) temperature measurement sensor, and (v) other simultaneous
heat transfer mechanisms such as convection and radiation. Some interesting
thought was exposed by~\citet{Nunes2003}, in order not to design
under or over-dimensioned heat exchangers or industrial equipments.
\citet{DiGuilio1992} discussed about the influence of electrical
charges, the pressure dependence, the presence of a saturation curve,
and the effect of size of ions in salt families. The method of measurement
is also a key issue for molten salts. This circumstance was also argued
by Diguilio and Teja and by Nieto de Castro, advising about examples
of positive or negative slope for the same salt by different authors.
In fact, Diguilio and Teja reported erroneous slope values for some
nitrates and chlorides.

Among correlation methods, many formulations have been developed and
reported with different assumptions, e.g., Bridgman and Kindcaid-Eyring
equation~\citep{Cooke1968}, Mean-ionic-weight and Rao-Turnbull correlations~\citep{Cornwell1971},
and Gustafsson and Rough-Hard-Sphere models~\citep{DiGuilio1992}.
\citet{Khokhlov2009} proposed a standard function of temperature
and molecular weight for multicomponent fluorides. In order to check
its behavior, some correlations have been extended for all the selected
salts in the present work.

\begin{figure*}
\centering{}\includegraphics[width=0.95\textwidth]{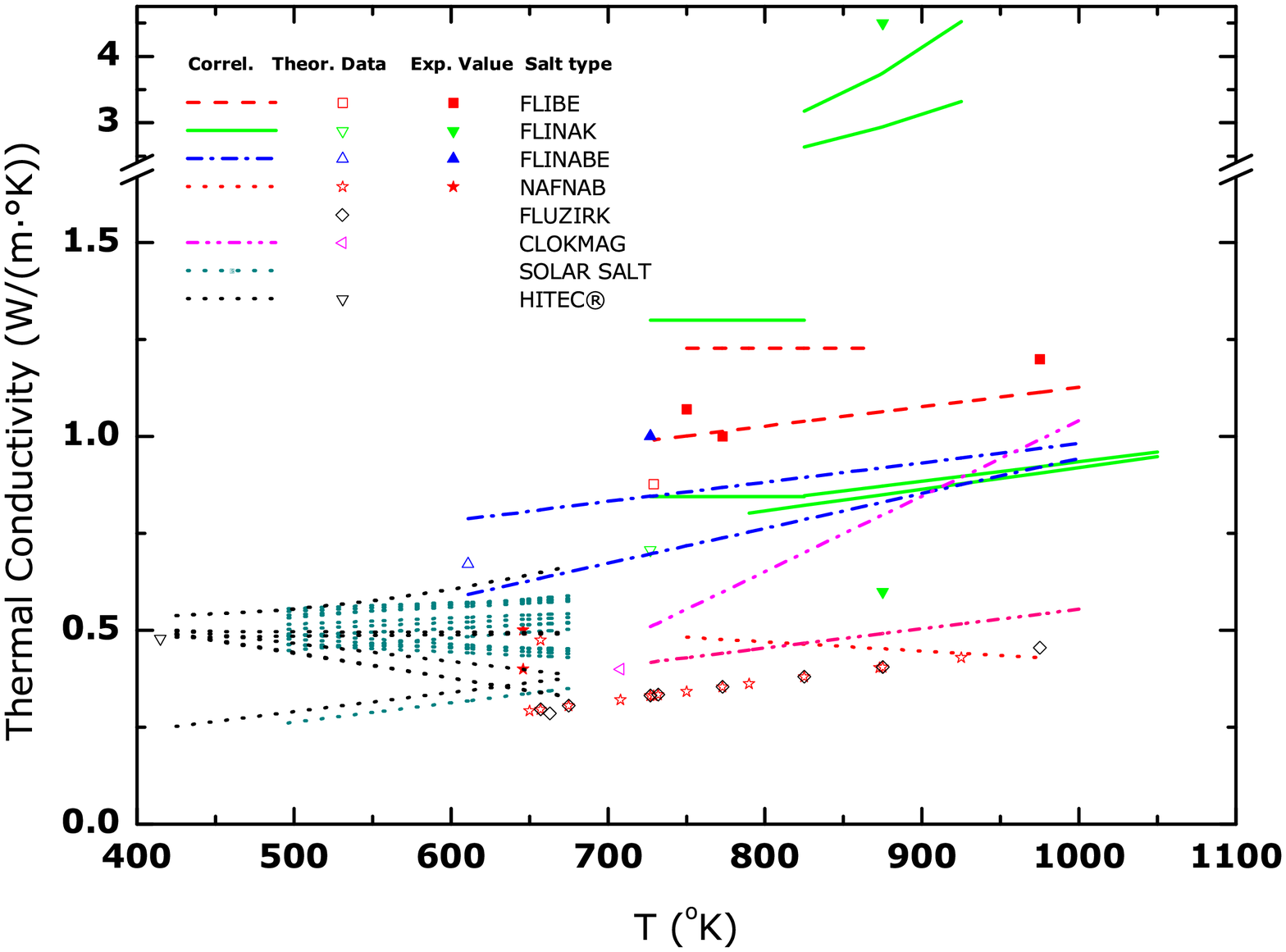}\caption{Graph summarizing the thermal conductivity correlations and estimated
values proposed by different authors at different temperatures. Legend
inside the graph shows: Lines defining different correlations reported
for each kind of salt; Symbols representing point values for theoretical
estimations (open symbols), or reported experimental measurements
(filled symbols).\label{fig:Sumarized-lambda}}
\end{figure*}

\noindent Both experimental or estimated data have been found in the
literature for the selected salt mixtures. After a global comparison,
a low temperature dependence is showed~\eqref{fig:Sumarized-lambda}.
Rao-Turnbull~\citep{Turnbull1961,Turnbull1961a} prediction shows
acceptable values for melting point, but is very sensitive to the
average number of ions per mole (a proposal of ions is listed in~\citet{Cooke1968}
for different mixtures). The assumptions for this correlation made
by~\citet{williams2006assess12} have been recalculated, obtaining
different values other than in this work. \citet{Khokhlov2009} functions
appears always with a positive slope, which is not true for some mixtures,
and the calculated values seem to be only coherent with measurements
for fluorides and chlorides (but not with fluoborates or nitrates). 

For FLiBe, we have plotted results and proposals from~\citet{Cantor_et_al1968,Cooke1968,Cooke1969,Kato1983,williams2006assess12}
and~\citet{Khokhlov2009}, and recalculated Rao-Turnbull correlation.
As the temperature dependence is relatively low, we agree with~\citet{Benes2012}
recommending $\mathrm{\lambda=1.1\, W\text{·}m^{-1}\cdot\text{\textdegree}K^{-1}}$,
with a 0.29 \% deviation from the average value.

FLiNaK investigations show a wide rage of data from 0.6 to 4.5 $\mathrm{W\cdot m^{-1}\cdot\text{\textdegree}K^{-1}}$.
The Rao-Turnbull value moves from 0.7 to 1.58 when the number of ions
is changed between 1 or 2. Several measurements and estimations have
been compared: \citet{grele1954forced,hoffman1955fused,lane1958fluid,Ewing1962,powers1963physical,vriesema1979aspects,janz1981physical,Kato1983,smirnov1987thermal}
and~\citet{Khokhlov2009}. Kato et al. values are derived from thermal
diffusivity data. Ewing et al. showed decreasing transmission coefficients,
which was explained due to solution of container components.

For the usual composition~\citet{Benes2012} conclusions recommended
a linear equation with positive slope. However,~\citet{DiGuilio1992}
discussed this behavior for the alkali halides (descent slope). In
any case, there is good agreement among~\citet{Kato1983} and~\citet{smirnov1987thermal}.
Therefore, a constant value of $\mathrm{\lambda=0.85\, W\text{·}m^{-1}\cdot\text{\textdegree}K^{-1}}$
is suggested in agreement with this two last authors and the expression
of~\citet{Khokhlov2009}. The standard deviation is too high if values
of~\citet{grele1954forced,hoffman1955fused,lane1958fluid,powers1963physical}
and~\citet{janz1981physical} are used in calculations, growing up
to 1.29; but it decreases till 0.21 when this anomalous values are
ignored. Using all data to compute the average the deviation of suggested
value is about 51.29 \%, but it only 10.78 \% when anomalous values
are not taken into account.

There are few estimations for FLiNaBe mixture, and most of them are
for different compositions. \citet{Grimes1967,Grimes1970,ignat2006experimental}
and~\citet{Khokhlov2009} have been revised in the present work.
Because of the lack of experimental data, a constant value of $\mathrm{\lambda=0.70\, W\text{·}m^{-1}\cdot\text{\textdegree}K^{-1}}$
is selected for the (0.31-0.31-0.28) mixture, which is coherent with
Rao-Turnbull correlation with n=1. The deviation from average is around
18.21 \% for the suggested heat capacity.

Several values have been published for NaFNaB with (0.08-0.92) molar
composition, such as~\citet{Cantor_et_al1968,Cooke1968,Cooke1969,Grimes1970}
and~\citet{Khokhlov2009}. Analyzing the proposed functions, thermal
conductivity can be evaluated by Rao-Turnbull equation (n=2), giving
$\mathrm{\lambda=0.47\, W\text{·}m^{-1}\cdot\text{\textdegree}K^{-1}}$
with 0.44 \% of deviation from average.

To our best knowledge, no measurements about FluZirK have been published.
A standard value, for (0.58-0.42) molar composition, can be estimated
in coherence with Rao-Turnbull expression (n=1) and Khokhlov et al.
correlation. Hence $\mathrm{\lambda=0.30\, W\text{·}m^{-1}\cdot\text{\textdegree}K^{-1}}$
is found to be a good constant value (19.61 \% of deviation from the
average).

The thermal conductivity of CloKMag have been reported by~\citet{janz1981physical}
for the (0.66-0.34) and (0.71-0.29) molar compositions. \citet{williams2006assess12}
calculated a value of 0.39 $\mathrm{W\cdot m^{-1}\cdot\text{\textdegree}K^{-1}}$
by simple mole-fraction average of the pure-compound data, which has
been recently suggested by~\citet{anderson2012molten}. By using
again the Rao-Turnbull estimation (n=2, n=1, and n=1.5), the values
obtained are 0.55, 0.25 and 0.40 $\mathrm{W\cdot m^{-1}\cdot\text{\textdegree}K^{-1}}$
respectively. The estimation function of~\citet{Khokhlov2009} is
near parallel to data obtained by~\citet{janz1981physical} correlation,
and agrees with Rao-Turnbull (n=1.5) and Williams et al. calculations.
Therefore, a constant value for thermal conductivity of $\mathrm{\lambda=0.55\, W\text{·}m^{-1}\cdot\text{\textdegree}K^{-1}}$
is suggested with temperature independence, in coherence with Rao-Turnbull
estimation for n=2. Deviation from average is about 14.02 \%, and
0.16 is the computed standard deviation.

The scattering of data is even larger for nitrate compositions, as
they have been studied extensively. Thermal decomposition of this
kind of salts must be taken also into account, which can be significantly
enhanced by controlling the atmosphere, as recently discussed by~\citet{Olivares2012}.
Several reports for Solar Salt (with different molar compositions)
have been reviewed, and also the SAM data have been correlated by
linear regression:~\citet{McDonald1970,janz1979physical,Omotani1982,Tufeu1985}
and~\citet{Zavoico2001}. According to~\citet{DiGuilio1992}, temperature
dependence must show negative slope. This feature is only followed
by Omotani et al. and Tufeu et al. expressions. \citet{Khokhlov2009}
function does not match the measured values. In general, correlated
values give a maximum of 0.58 and a minimum of 0.42 $\mathrm{W\text{·}m^{-1}\cdot\text{\textdegree}K^{-1}}$
for the range $\mathrm{T\in[600-730]}$, and Rao-Turnbull gives 0.47
$\mathrm{W\text{·}m^{-1}\cdot\text{\textdegree}K^{-1}}$. The latter
value agrees with Tufeu et al. and Omotani et al. correlations, and~\citet{DiGuilio1992}
predictions. Hence, a constant $\mathrm{\lambda=0.45\, W\text{·}m^{-1}\cdot\text{\textdegree}K^{-1}}$
is found to be a good choice, showing a 10.12 \% of deviation from
average.

Finally, Hitec$^{\lyxmathsym{\textregistered}}$ correlations have
been reviewed. SAM database has been also accessed, in order to compare
values with~\citet{Cooke1968,janz1981physical,Omotani1984,Santini1984,Tufeu1985,Yang2010,Wu2012}
and~\citet{Boerema2012}. The plotted functions of temperature show
a dispersion around an average value, which is close to Rao-Turnbull
estimation for n=2 and giving $\mathrm{\lambda=0.48\, W\text{·}m^{-1}\cdot\text{\textdegree}K^{-1}}$.
The latter is also very close to Tufeu et al. and Omotani et al. values.
The global standard deviation for this property and salt has been
computed about 0.097. Although there is a general disagreement among
most recent reports, a suggested value of $\mathrm{\lambda=0.48\, W\text{·}m^{-1}\cdot\text{\textdegree}K^{-1}}$
is also in coherence with~\citet{DiGuilio1992} arguments for $\mathrm{KNO_{3}}$
and nitrate mixtures. The deviation from average for this last value
is calculated to be around 4.36 \%.

\noindent 

Trying to make a brief summary for this property,~\ref{fig:Sumarized-lambda}
shows a global view of reviewed values for thermal conductivity and
all studied salts. Although most of mixtures shows a high scattering
when studied separately, some conclusions can be made with a general
overview. For example, most of data are bounded into the range $\mathrm{0.25\,-\,1.30\, W\text{·}m^{-1}\cdot\text{\textdegree}K^{-1}}$
. In case of nitrate-nitrite salts, this interval can be reduced to
$\mathrm{0.25\,-\,0.70\, W\text{·}m^{-1}\cdot\text{\textdegree}K^{-1}}$,
with independence of temperature value. For fluorides, data are bounded
into $\mathrm{0.25\,-\,1.30\, W\text{·}m^{-1}\cdot\text{\textdegree}K^{-1}}$
if anomalous values reported by~\citet{grele1954forced,hoffman1955fused}
and~\citet{janz1981physical} for FLiNaK are ignored. In general,
more tests are needed so as to asses better or newer correlations
for salts at high temperature conditions. Hence, with currently knowledge,
a constant value may be a good option to get acceptable accuracy in
calculations~(\ref{tab:th_cond_values}).

\noindent 
\begin{table}[h]
\caption{\label{tab:th_cond_values}\protect \\
Summary of suggested constant values for thermal conductivity~$\mathrm{(\lambda,\, W\cdot m^{-1}\cdot\text{\textdegree}K^{-1}})$,
with temperature independence. Table include the calculated standard
deviation for studied data, and \% of deviation of selected values
from the average. Deviation from average for FLiNaK is too high if
anomalous high values are not avoided in calculations (51.29 \%),
but decrease to 10.78 \% when this values are not used.}

\noindent \centering{}%
\begin{tabular}{cccc}
\toprule 
{\scriptsize Salt mixture} & {\scriptsize Constant value} & {\scriptsize Stand. Dev.} & {\scriptsize \% Dev.}\tabularnewline
\midrule
\midrule 
{\scriptsize FLiBe} & {\scriptsize 1.10} & {\scriptsize 0.096} & {\scriptsize 0.29 \%}\tabularnewline
\midrule 
{\scriptsize FLiNaK} & {\scriptsize 0.85} & {\scriptsize 0.206} & {\scriptsize 10.78 \%}\tabularnewline
\midrule 
{\scriptsize FLiNaBe} & {\scriptsize 0.70} & {\scriptsize 0.105} & {\scriptsize 18.21 \%}\tabularnewline
\midrule 
{\scriptsize NaFNaB} & {\scriptsize 0.47} & {\scriptsize 0.016} & {\scriptsize 0.44 \%}\tabularnewline
\midrule 
{\scriptsize FluZirK} & {\scriptsize 0.30} & {\scriptsize 0.054} & {\scriptsize 19.61 \%}\tabularnewline
\midrule 
{\scriptsize CloKMag} & {\scriptsize 0.55} & {\scriptsize 0.163} & {\scriptsize 14.02 \%}\tabularnewline
\midrule 
{\scriptsize Solar Salt} & {\scriptsize 0.45} & {\scriptsize 0.067} & {\scriptsize 10.12 \%}\tabularnewline
\midrule 
{\scriptsize Hitec$^{\lyxmathsym{\textregistered}}$} & {\scriptsize 0.48} & {\scriptsize 0.097} & {\scriptsize 4.36 \%}\tabularnewline
\bottomrule
\end{tabular}
\end{table}

\subsection{Heat capacity (\emph{Cp},~J·$\mathrm{kg^{-1}}$·~\textmd{\textdegree{}}$\mathrm{K^{-1}}$)}

\begin{figure}
\centering{}\includegraphics[width=8.5cm]{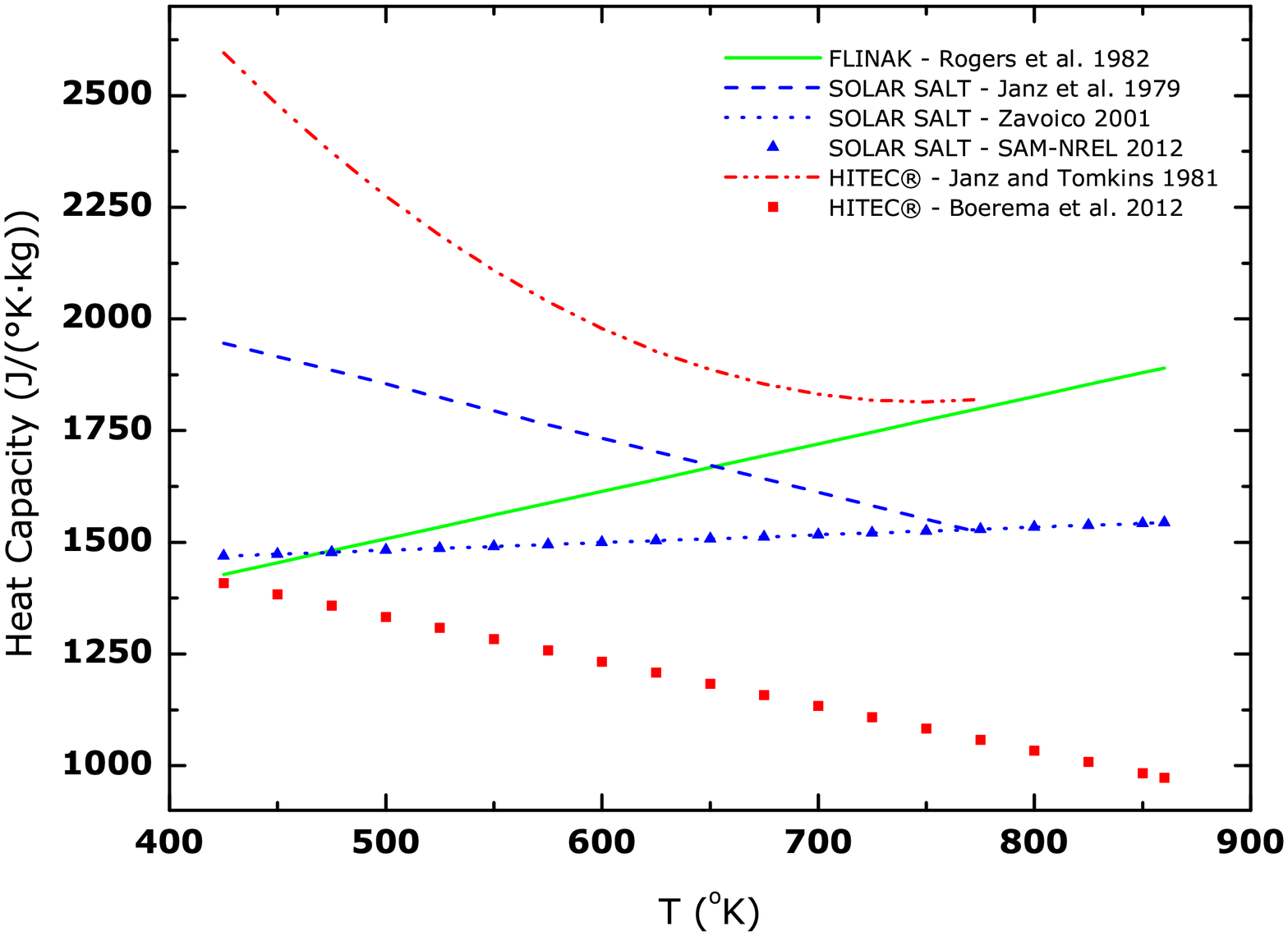}\caption{Correlations found for heat capacity in the case of FLiNaK~(\textcolor{green}{\scriptsize --}),
Solar Salt~(\textcolor{blue}{\scriptsize - - -}, \textcolor{blue}{\scriptsize ·~·~·
},\textcolor{blue}{\scriptsize{} ${\color{blue}{\normalcolor {\color{blue}\blacktriangle}}}$})
and Hitec$^{\lyxmathsym{\textregistered}}$~(\textcolor{red}{\scriptsize -~·~-},
\textcolor{red}{\tiny $\blacksquare$}) mixtures and usual compositions.\label{fig:Cp_correl} }
\end{figure}

\begin{figure*}
\centering{}\includegraphics[width=0.95\linewidth]{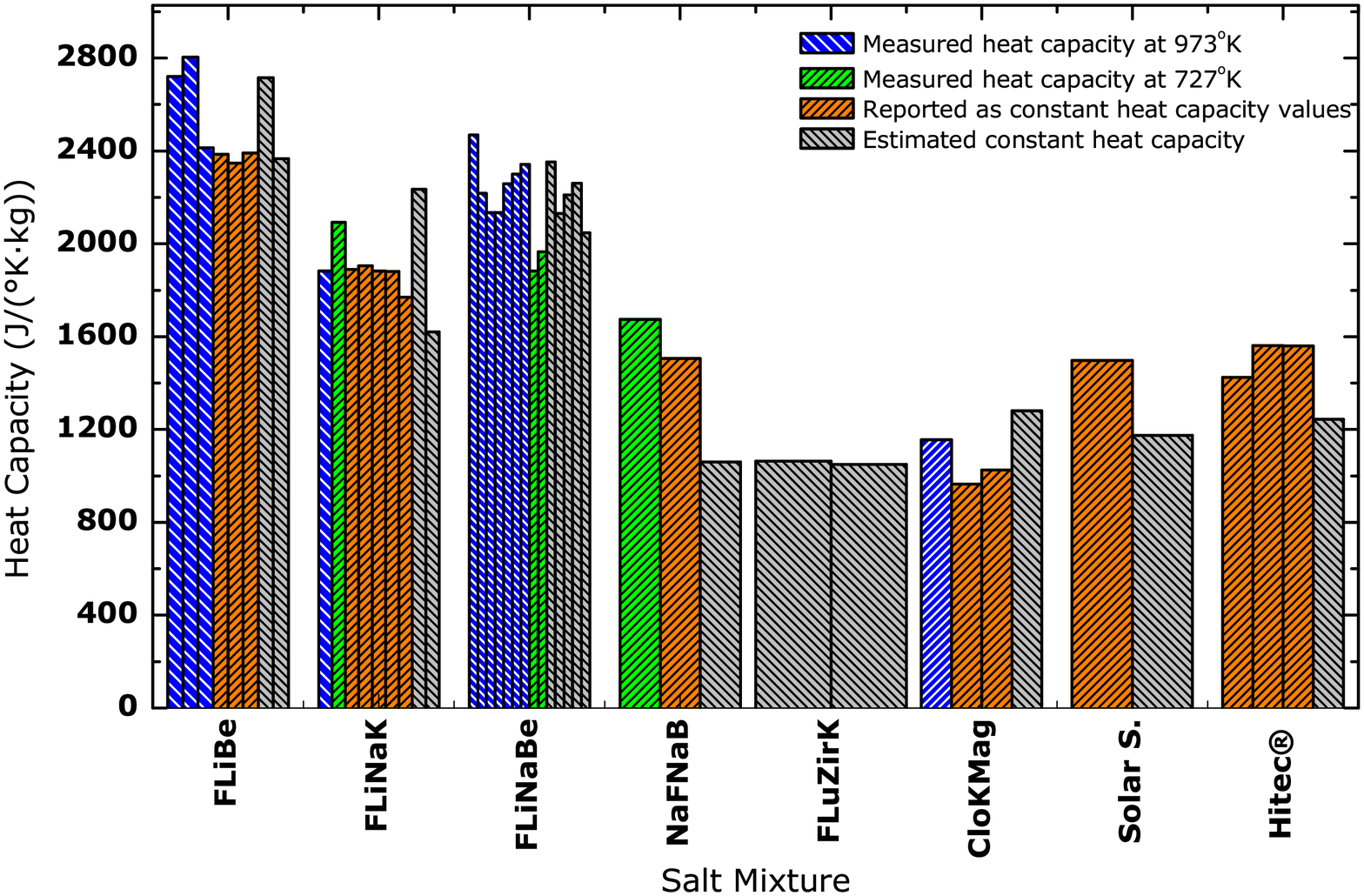}\caption{Global comparison of reported values of heat capacity for selected
salts and different compositions.\label{fig:Global_Cp}}
\end{figure*}

\noindent \citet{Khokhlov2009}~reported a general correlation for
multi-component fluorides. In order to asses further possibilities,
this expression have been made extensive to the other molten salts
in this work. Some other functions of temperature have been found
for certain salts as plotted in~\ref{fig:Cp_correl}. \citet{Cantor1965}
proposed the assumption of temperature independence for this property,
because of the small accuracy observed in experiments. Cantor modified
the Dulong-Petit expression for molten fluorides, using an average
value for heat capacity per atom (33.472 J/\textdegree{}K). This proposal
was used by~\citet{williams2006assess12}, making a comparison with
measurements at 973 \textdegree{}K. A global graph is showed in \ref{fig:Global_Cp},
including reported and estimated values. The suggested values for
each salt are summarized in~\ref{tab:Cp_values}.

For FLiBe, values were reported by~\citet{lane1958fluid} for (0.69-0.31)
and (0.50-0.50) compositions at 973 \textdegree{}K. \citet{Cantor_et_al1968}
(from an internal report of Hoffman and Lones) and~\citet{douglas1969measured}
also gave values for the promising (0.66-0.34) mixture. Heat capacity
of FLiBe can be also obtained of~\citet{Kato1983} from thermal diffusivity.
By comparison, we propose~\citet{Cantor_et_al1968} as a constant
value, giving $\mathrm{\mathit{Cp}=2385\, J\cdot kg^{-1}\cdot\text{\textdegree}K^{-1}}$
and a 5.26 \% of deviation from the average.

\noindent 

\noindent 

First found data for FLiNaK were published by~\citet{poppendick1952physical},
and also by Powers and Blalock at ORNL~\citep{Powers1953} using
a Bunsen calorimeter. \citet{grele1954forced} used 2092 J·kg$^{-1}$·\textdegree{}K$^{-1}$
with temperature independence, based on data reported by Poppendick.
Other proposals were made by~\citet{janz1981physical,Kato1983} and~\citet{Salanne2009_26}
(this last, from a first-principles determination). Kato et al. values
are based on thermal diffusivity data. \citet{vriesema1979aspects}
used a constant value of 1890 J·kg$^{-1}$·\textdegree{}K$^{-1}$.
\citet{rogers1982fusion} also reported a correlation for this property
with positive slope, as showed in~\ref{fig:Cp_correl}. Published
data are higher than ideal behavior in agreement with~\citet{Benes2012}.
After verifying the strong correlation among the constant reported
data (standard deviation of 208.46), a heat capacity of $\mathrm{\mathit{Cp}=1880\, J\cdot kg^{-1}\cdot\text{\textdegree}K^{-1}}$
is suggested to be a good value with 6.07 \% of deviation from average.

For the molten FLiNaBe, several values have been reported for different
molar compositions. However, none of them for the equimolar or (0.31-0.31-0.38)
compositions. \citet{lane1958fluid} gave 2467 J·kg$^{-1}$·\textdegree{}K$^{-1}$
value for (0.35-0.27-0.38) at 973\textdegree{}K, and~\citet{powers1963physical}
reviewed for different molar percentages at the same temperature,
as well as~\citet{Grimes1967} at 727 \textdegree{}K. A global standard
deviation of 158.15 has been calculated for reviewed data. For the
selected composition (0.31-0.31-0.38), we suggest a constant heat
capacity of $\mathrm{\mathit{Cp}=2200\, J\cdot kg^{-1}\cdot\text{\textdegree}K^{-1}}$
which has a 0.48 \% of deviation from average.

No measurements have been found for FluZirK. Only estimated values
can be obtained by the mentioned~\citet{Khokhlov2009} and Cantor-Dulong-Petit
correlations. A constant heat capacity of $\mathrm{\mathit{Cp}=1000\, J\cdot kg^{-1}\cdot\text{\textdegree}K^{-1}}$
is prudently proposed with 5.36 \% of deviation from the average of
estimations.

Only two references have been found for NaFNaB (0.08-0.92). \citet{Grimes1967}
reported a value of 1674 J·kg$^{-1}$·\textdegree{}K$^{-1}$, and
as mentioned in~\citet{Cantor_et_al1968}, Dworkin measured it giving
$\mathrm{\mathit{Cp}=1506\, J\cdot kg^{-1}\cdot\text{\textdegree}K^{-1}}$,
which is the suggested value for this salt. Calculated deviation from
the average is about 6.56 \% in this case.

For CloKMag (0.666-0.334) and (0.418-0.582) compositions,~\citet{janz1981physical}
reported 964 and 1026 J·kg$^{-1}$·\textdegree{}K$^{-1}$ respectively.
\citet{williams2006assess69} also reported a measured value at 973
\textdegree{}K of $\mathrm{\mathit{Cp}=1155\, J\cdot kg^{-1}\cdot\text{\textdegree}K^{-1}}$for
the mixture (0.67-0.33), which is suggested as good value (4.42 \%
of deviation from average).

The nitrate and nitrite salts, in addition to the correlations plotted
in~\ref{fig:Cp_correl}, have been reported several times elsewhere
(e.g.,~\citet{janz1979physical,janz1981physical}, among others).
The heat capacity of equimolar Solar Salt was fixed in 1498 J·kg$^{-1}$·\textdegree{}K$^{-1}$
by~\citet{Tufeu1985}. Although the correlated SAM data~\citet{NREL}
and the expression reported by~\citet{Zavoico2001} are exactly the
same, some differences can be found for temperature dependence if
compared with~\citet{janz1979physical}. Agreement among the different
investigations analyzed (standard deviation about 148.99) suggests
that the actualized values offered by SAM may be a good choice with
a 2.36 \% of deviation from the global average, using the following
expression:

\begin{equation}
Cp\,\mathrm{(J\cdot kg^{-1}\cdot\text{\textdegree}K^{-1})=1396.044+0.172\text{·}T(\text{\textdegree}K)}\label{eq:CP_solar-salt}
\end{equation}

Finally, correlations for the commercial Hitec$^{\lyxmathsym{\textregistered}}$
have been reported by~\citet{Hoffman1960,janz1981physical} and~\citet{Boerema2012},
using different expressions. Constant values were reported by~\citet{Wu2012}
and~\citet{Yang2010}, and this latter gives the same number as found
in the SAM database. The computed standard deviation is 473.23 when
using all reviewed data. A value of $\mathrm{\mathit{Cp}=1560\, J\cdot kg^{-1}\cdot\text{\textdegree}K^{-1}}$
is suggested due to the agreement of recent proposals with early measurements,
giving a 2.45 \% of deviation from average.

\noindent 
\begin{table}
\caption{\label{tab:Cp_values}\protect \\
Heat capacity~$\mathrm{(J\cdot kg^{-1}\cdot\text{\textdegree}K^{-1})}$,
suggested as constant value with temperature independence, for the
reviewed salts. A function is only offered for the Solar Salt mixture.}

\noindent \centering{}%
\begin{tabular}{cccc}
\toprule 
{\scriptsize Salt mixture} & \emph{\scriptsize Cp}{\scriptsize{} value} & {\scriptsize Stand. Dev.} & {\scriptsize \% Dev.}\tabularnewline
\midrule
\midrule 
{\scriptsize FLiBe} & {\scriptsize 2385} & {\scriptsize 191.89} & {\scriptsize 5.26 \%}\tabularnewline
\midrule 
{\scriptsize FLiNaK} & {\scriptsize 1880} & {\scriptsize 208.46} & {\scriptsize 6.07 \%}\tabularnewline
\midrule 
{\scriptsize FLiNaBe} & {\scriptsize 2200} & {\scriptsize 158.15} & {\scriptsize 0.48 \%}\tabularnewline
\midrule 
{\scriptsize NaFNaB} & {\scriptsize 1506} & {\scriptsize 263.08} & {\scriptsize 6.56 \%}\tabularnewline
\midrule 
{\scriptsize FluZirK} & {\scriptsize 1000} & {\scriptsize 9.43} & {\scriptsize 5.36 \%}\tabularnewline
\midrule 
{\scriptsize CloKMag} & {\scriptsize 1155} & {\scriptsize 140.41} & {\scriptsize 4.42 \%}\tabularnewline
\midrule 
{\scriptsize Solar Salt} & {\scriptsize 1396.044+0.172·T} & {\scriptsize 148.99} & {\scriptsize 2.36 \%}\tabularnewline
\midrule 
{\scriptsize Hitec$^{\lyxmathsym{\textregistered}}$} & {\scriptsize 1560} & {\scriptsize 473.23} & {\scriptsize 2.45 \%}\tabularnewline
\bottomrule
\end{tabular}
\end{table}

\section{Conclusions}

The future use of molten salts as coolants or HTF requires a previous
checking of transport and thermal behaviors. The use of computational
packages allows validation of preliminary designs, and even complete
piping systems, reactors, or heat exchanger loops. Moreover, numerical
simulations need verified physical properties as an input. This work
intends to be a refined compendium of data, aiming to feed computer
aided engineering designs. After a intensive review of the different
reports and published data for eight different mixtures, several discrepancies
have been found for some correlations, and a lack of data for certain
salts. Deviations reveal that density is the best known property,
but both viscosity as values for thermal properties, which in some
cases are only based in theoretical models, show scattering. By this
order, FLiBe, nitrates and NaFNaB mixtures have a global acceptable
accuracy (less than 5 \% of global average deviations), followed by
CloKMag, FLiNaK and FLiNaBe (less than 8 \%). Although FluZirK has
been recently suggested as suitable salt for certain applications,
the level of knowledge about transport and thermal behavior is very
short nowadays. Hence, the need of new studies is legitimated in order
to obtain higher accuracy, mainly for thermal properties. Discussion
about this latter parameters have shown that standard techniques must
be refined or even developed in some particular cases, to avoid undesirable
conditions or mechanisms which can disturb measuring procedures (such
as interaction of sensors or dissolution of container materials).

\bibliographystyle{model1a-num-names}
\bibliography{library_final}

\end{document}